\documentstyle[12pt,graphicx]{article}
\begin{document}
\baselineskip 0.6cm

\def\simgt{\mathrel{\lower2.5pt\vbox{\lineskip=0pt\baselineskip=0pt
           \hbox{$>$}\hbox{$\sim$}}}}
\def\simlt{\mathrel{\lower2.5pt\vbox{\lineskip=0pt\baselineskip=0pt
           \hbox{$<$}\hbox{$\sim$}}}}

\begin{titlepage}

\begin{flushright}
SLAC-PUB-11892 \\
UCB-PTH-06/11 \\
LBNL-60363
\end{flushright}

\vskip 1.7cm

\begin{center}

{\Large \bf 
Supersymmetry with Small {\boldmath $\mu$}: Connections between \\
Naturalness, Dark Matter, and (Possibly) Flavor
}

\vskip 1.0cm

{\large
Ryuichiro Kitano$^a$ and Yasunori Nomura$^{b,c}$}

\vskip 0.4cm

$^a$ {\it Stanford Linear Accelerator Center, 
                Stanford University, Stanford, CA 94309} \\
$^b$ {\it Department of Physics, University of California,
                Berkeley, CA 94720} \\
$^c$ {\it Theoretical Physics Group, Lawrence Berkeley National Laboratory,
                Berkeley, CA 94720} \\

\vskip 1.2cm

\abstract{Weak scale supersymmetric theories often suffer from several 
naturalness problems: the problems of reproducing the correct scale for 
electroweak symmetry breaking, the correct abundance for dark matter, 
and small rates for flavor violating processes.  We argue that the 
first two problems point to particular regions of parameter space in 
models with weak scale supersymmetry: those with a small $\mu$ term. 
This has an interesting implication on direct dark matter detection 
experiments.  We find that, if the signs of the three gaugino mass 
parameters are all equal, we can obtain a solid lower bound on the 
spin-independent neutralino-nucleon cross section, $\sigma_{\rm SI}$. 
In the case that the gaugino masses satisfy the unified mass relations, 
we obtain $\sigma_{\rm SI} \simgt 4 \times 10^{-46}~{\rm cm}^2$ ($1 
\times 10^{-46}~{\rm cm}^2$) for fine-tuning in electroweak symmetry 
breaking no worse than $10\%$ ($5\%$).  We also discuss a possibility 
that the three problems listed above are all connected to the hierarchy 
of fermion masses.  This occurs if supersymmetry breaking and electroweak 
symmetry breaking (the Higgs fields) are coupled to matter fields with 
similar hierarchical structures.  The discovery of $\mu \to e$ transition 
processes in near future experiments is predicted in such a framework.}

\end{center}
\end{titlepage}

\section{Introduction}

Weak scale supersymmetry provides an elegant framework to solve 
the naturalness problem of the standard model as well as to explain 
the dark matter of the universe.  An extreme sensitivity of the weak 
scale to ultraviolet physics in the standard model is softened due to 
the existence of superparticles at this scale, and the stable lightest 
supersymmetric particle (LSP) left over from the early history of the 
universe composes dark matter today.  These qualitative successes, 
however, should now be reviewed much more carefully.  On one hand, 
non-discovery of both superparticles and a light Higgs boson at 
LEP~II~\cite{Barate:2003sz} raises the overall mass scale for the 
superparticles, leading to a tension with naturalness of electroweak 
symmetry breaking.  On the other hand, the accurate measurement of 
the dark matter density by WMAP~\cite{Spergel:2003cb} gives a precise 
constraint on the spectrum of superparticles.  It is plausible that 
a careful study of these issues provides strong hints on a possible 
realization of supersymmetry at the weak scale.

Phenomenology of supersymmetric theories depends strongly on how 
fundamental supersymmetry breaking is mediated to the supersymmetric 
standard model sector.  What is the underlying mechanism of the mediation? 
A promising possibility arises if mediation occurs through gravitationally 
suppressed interactions.  This has a virtue that the supersymmetric 
mass term for the Higgs doublets ($\mu$ term) is naturally generated 
with the weak scale size, because it can arise as a sort of supersymmetry 
breaking term in this case~\cite{Giudice:1988yz}.  This provides an 
elegant ``solution'' to the $\mu$ problem, which plagues many other 
scenarios for supersymmetry breaking.  Another virtue of mediation 
by gravitational strength interactions is that the gravitino is likely 
to be heavier than the lightest neutralino, giving a possibility 
of weakly interacting massive particle (WIMP) dark matter.  Such 
a mediation is also minimal in the sense that it does not require 
any other physics than that at the gravitational scale, which we know 
exists.  We thus mainly focus on this class of mediation -- gravity 
mediation broadly defined -- in this paper, and study what current 
experimental data imply on the structure of superparticle spectra. 
We argue that the current data strongly suggest that the $\mu$ term 
is small, $|\mu| \simlt (200\!\sim\!400)~{\rm GeV}$, regardless of 
any details of supersymmetry breaking.  This in turn has a striking 
consequence in the context of the minimal supersymmetric standard 
model (MSSM) --- the cross section for direct dark matter detection 
is large and can naturally be in the range where the CDMS~II experiment 
will explore in the next two years.  In fact, we can obtain a solid 
lower bound on the cross section if the signs of the gaugino masses 
are universal.   In the case that the gaugino masses satisfy the unified 
mass relations, we obtain $\sigma \simgt 4 \times 10^{-46}~{\rm cm}^2$ 
($1 \times 10^{-46}~{\rm cm}^2$) for fine-tuning in electroweak 
symmetry breaking no worse than $10\%$ ($5\%$).

An important consequence of mediation by gravitational scale 
interactions is that it opens up a window to connect physics at the 
weak scale to that at high energies, such as the Planck or unification 
scale, since the low energy scalar (squark, slepton and Higgs boson) 
masses carry all information from the Planck to the weak scales through 
their renormalization group evolutions.  The possibility of perturbative 
extrapolations of physics across a vast energy interval, in fact, is 
a unique feature of theories with weak scale supersymmetry and supported 
by the successful unification of gauge couplings at a scale of $\approx 
10^{16}~{\rm GeV}$~\cite{Dimopoulos:1981zb}.  We thus explore possible 
implications of low energy spectra suggested by the current data 
(small $\mu$ term) on physics at the gravitational scale.  We find 
that a desired patten of superparticle spectra is obtained if physics 
of supersymmetry breaking mediation is intimately related to that of 
flavor.  This has interesting implications on low energy experiments 
exploring flavor violation.  We present an explicit scheme incorporating 
these ideas, which we call next to minimal supergravity, and estimate 
rates of various flavor violating processes in this framework.  We 
find that some of these processes, such as $\mu \rightarrow e\gamma$, 
are naturally close to the current experimental bounds, so that they 
are expected to be within the reach of near future experiments.

The organization of the paper is as follows.  In section~\ref{sec:DM} 
we consider a connection between naturalness in electroweak symmetry 
breaking and physics of dark matter.  We see that a small value for 
the $\mu$ parameter, required from naturalness of electroweak symmetry 
breaking, leads quite naturally to a thermal relic abundance of the 
lightest neutralino consistent with the WMAP data.  We also perform 
a ``model independent'' analysis of the direct detection cross section 
for such dark matter, and find that it is generically large.  In 
section~\ref{sec:model} we explore possible high energy theories that 
lead to superparticle spectra identified in section~\ref{sec:DM}.  
As one of such theories, we consider a scenario in which physics of 
supersymmetry breaking is intimately related to that of flavor.  We 
estimate various flavor violating processes, and find that lepton 
flavor violating processes in the first two generations are generically 
large and close to the current experimental bounds.  Finally, conclusions 
are given in section~\ref{sec:concl}.

\section{Small {\boldmath $\mu$} Term: Connection between Electroweak 
 Naturalness and Dark Matter}
\label{sec:DM}

In this section we see a close connection between naturalness of 
electroweak symmetry breaking and physics of dark matter, suggested 
by the data from LEP~II and WMAP.  These data both seem to suggest 
a certain parameter space in weak scale supersymmetry.  We find that 
this has an interesting consequence on the detection of dark matter 
in the context of the MSSM.

Let us begin our discussion by reviewing the situation in minimal 
supergravity (mSUGRA)~\cite{Chamseddine:1982jx}: a popular scenario for 
gravitational mediation, in which certain flavor universal interactions 
between the supersymmetry breaking sector and the supersymmetric 
standard model sector are assumed to be responsible for the mediation. 
This scenario provides a simple parameterization of relevant physics 
at the Planck scale.  In the simplest case, the supersymmetry breaking 
masses at the gravitational/unification scale are parameterized 
by five free parameters: the universal gaugino mass $M_{1/2}$, the 
universal scalar squared mass $m_0^2$, the universal scalar trilinear 
interaction $A_0$, the supersymmetric Higgs mass $\mu$, and the 
holomorphic supersymmetry breaking Higgs squared mass $\mu B$. 
While this setup leads to qualitatively correct low energy physics, 
it has become gradually clearer that it does not seem to give a very 
good description of our world at the quantitative level.  In particular, 
given the current experimental constraints on the superparticle and 
the Higgs boson masses, a typical parameter region of mSUGRA leads to 
too large electroweak symmetry breaking and too large relic abundance 
for the dark matter.  The five independent parameters must be finely 
tuned to reproduce the observed electroweak symmetry breaking scale 
as well as the correct amount of the dark matter, determined precisely 
by the recent WMAP data~\cite{Spergel:2003cb}.  In view of these 
unpleasant situations, it seems clear that we must deviate from 
the simplest mSUGRA scenario to account for latest observations 
in a natural manner.

What direction should we take?  Looking at carefully the problems 
of the simplest mSUGRA described above, we find that these seemingly 
unrelated problems are in fact somewhat correlated.  Let us first 
consider the issue of electroweak symmetry breaking.  For reasonably 
large values of $\tan\beta \equiv \langle H_u \rangle/\langle H_d 
\rangle$, e.g. $\tan\beta \simgt 3$, which is suggested by the LEP~II 
lower bound on the Higgs boson mass~\cite{Barate:2003sz}, the equation 
determining the electroweak symmetry breaking scale is given by
\begin{equation}
  \frac{M_{\rm Higgs}^2}{2} \simeq - m_{H_u}^2 - |\mu|^2,
\label{eq:ewsb}
\end{equation}
where $M_{\rm Higgs}$ represents the mass of the lightest $CP$-even 
Higgs boson, and $m_{H_u}^2$ the soft supersymmetry breaking squared 
mass for the up-type Higgs boson $H_u$ ($m_{H_u}^2 < 0$).  In the 
MSSM, $M_{\rm Higgs}$ cannot be larger than $\simeq 120~{\rm GeV}$ 
without significant fine-tuning of parameters.  This implies that 
each term in the right-hand-side of Eq.~(\ref{eq:ewsb}) cannot 
be much larger than $\simeq (90~{\rm GeV})^2$ if we want to avoid 
fine-tuning between two independent parameters $m_{H_u}^2$ and $\mu$. 
In fact, this is not so straightforward to achieve, because $m_{H_u}^2$ 
generically receives large radiative corrections from top-stop loops, 
which are logarithmically enhanced for the case of gravity mediation. 
Indeed, in the simplest mSUGRA, $|m_{H_u}^2|$ is generically quite 
large at the weak scale, leading to fine-tuning in electroweak symmetry 
breaking.

While lowering $|m_{H_u}^2|$ is an issue, which has been one of 
the main focuses in efforts trying to reduce fine-tuning, there is 
a general consequence of Eq.~(\ref{eq:ewsb}) which applies to any 
theories that do not extend the Higgs sector drastically at the 
weak scale --- in order for a theory to be natural, the $\mu$ term 
should not be larger than $100~{\rm GeV}$ by more than a factor 
of a few {\it no matter what the mechanism of reducing $|m_{H_u}^2|$ 
is}.  Requiring that the cancellation between the two terms in 
Eq.~(\ref{eq:ewsb}) is no worse than $\Delta^{-1}$, we obtain 
\begin{equation}
  |\mu| \simlt (270~{\rm GeV})
    \Biggl(\frac{10\%}{\Delta^{-1}}\Biggr)^{1/2},
\label{eq:mu-bound}
\end{equation}
for $M_{\rm Higgs} \simeq 120~{\rm GeV}$.  This has a striking 
consequence on physics of dark matter.  Since the lightest neutralino, 
which is assumed to be the LSP, contains a non-negligible mixture of 
the Higgsino, the relic abundance is reduced compared with a typical 
mSUGRA parameter region, reproducing the observed dark matter abundance 
quite naturally.  A large Higgsino fraction in the lightest neutralino 
also dramatically increases the possibility of detecting the dark 
matter, as we will see shortly.  A connection between naturalness 
of electroweak symmetry breaking and the detectability of dark matter 
has been emphasized in Ref.~\cite{Kitano:2005ew} in the context 
of a model solving the fine-tuning problem (for earlier work, 
see e.g.~\cite{Corsetti:2000yq}), and neutralino dark matter with 
non-negligible Higgsino mixture has been studied recently in various 
contexts, e.g., in~[\ref{Ellis:2002wv:X}~--~\ref{Baer:2006dz:X}] 
(see e.g.~[\ref{Berezinsky:1995cj:X}~--~\ref{Drees:2000bs:X}] 
for earlier work).%
\footnote{In fact, smallness of $|\mu|$ should generally be true 
in any theory (not necessarily the MSSM) which does not have severe 
fine-tuning.  For example, this is true for $\mu B$-driven electroweak 
symmetry breaking discussed in Ref.~\cite{Nomura:2005rk}, where 
the equation determining the electroweak scale differs from that 
in Eq.~(\ref{eq:ewsb}).  (The bound in this case is somewhat 
weaker and is given by $|\mu| \simlt \lambda\, (390~{\rm GeV}) 
(10\%/\Delta^{-1})^{1/2}$, where $v \simeq 174~{\rm GeV}$ and $\lambda 
\simlt (2\!\sim\!3)$ in general.)  If there is a singlet field whose 
vacuum expectation value (VEV) contributes to the Higgsino mass, 
we should simply replace $|\mu|$ by $|\mu_{\rm eff}|$ involving the 
singlet VEV.  Implications on dark matter physics, discussed in this 
paper, then mostly persist in these extended models, unless the LSP 
contains a significant amount of the singlino.  In particular, the 
correct thermal relic abundance is naturally obtained for small 
$\mu_{\rm eff}$, since it does not depend very strongly on the value 
of $\tan\beta$.}

The argument described above suggests that a key point of making 
supersymmetric theories natural is to lower the value of $\mu$, 
and thus the value of $|m_{H_u}^2|$.  This has the following far 
reaching implications in the context of minimal supersymmetric 
models~\cite{Kitano:2006gv,Kitano:2005wc}.  First, barring the 
possibility of accidental cancellations, small values of $|m_{H_u}^2|$ 
imply that the top squarks should be light because $|m_{H_u}^2|$ receives 
radiative corrections proportional to the squared masses of the top 
squarks, $m_{\tilde{t}}^2$, through the $O(1)$ top Yukawa coupling. 
Then, to satisfy the LEP~II bound on the physical Higgs boson mass 
$M_{\rm Higgs} \simgt 114~{\rm GeV}$, the scalar trilinear interaction 
for the top squarks, $A_t$, should be large at the weak scale:
\begin{equation}
  \left|\frac{A_t}{m_{\tilde{t}}}\right| 
  \approx (1.5\!\sim\!2.5).
\label{eq:At}
\end{equation}
In gravity mediation, this leads to the following two requirements 
on the soft supersymmetry breaking parameters at the unification 
scale $M_U$: the value of $A_t$ should be non-zero and negative, 
$A_t < 0$, and the gluino mass $M_3$ should be reasonably large, 
$M_3 \simgt 150~{\rm GeV}$.  (Throughout the paper, our sign 
convention for $\mu$ and the soft supersymmetry breaking parameters 
follows that of SUSY Les Houches Accord~\cite{Skands:2003cj}. 
For the case of non-universal gaugino masses, we take the convention 
that the gluino mass parameter, $M_3$, is positive.)  These 
requirements come from the fact that we cannot obtain large enough 
$|A_t/m_{\tilde{t}}|$ at the weak scale for $A_t(M_U) = 0$, and yet 
the sensitivity of low energy $A_t$ to $A_t(M_U)$ is so weak that 
we also need a renormalization group contribution to $A_t$ from 
$M_3$, which drives $A_t$ negative at the infrared.  This has 
a consequence that
\begin{equation}
  A_t < 0,
\label{eq:sgn-At}
\end{equation}
at the weak scale.  Another important constraint on the top squark 
sector is that the top squarks should be light.  From consideration 
of infrared contribution to $m_{H_u}^2$ from $m_{\tilde{t}}^2$ 
alone, we find that $m_{\tilde{t}}^2$ should not be much larger 
than $\approx (300~{\rm GeV})^2$.  With these light top squarks, 
the lightest Higgs boson can evade the LEP~II bound only if the 
tree-level contribution is sizable.  In the MSSM, this leads to 
the bound
\begin{equation}
  \tan\beta \simgt 5.
\label{eq:tan-beta}
\end{equation}
Note that the conditions of Eqs.~(\ref{eq:At}~--~\ref{eq:tan-beta}) 
should be satisfied for any minimal and natural supersymmetric theories 
in which supersymmetry breaking is mediated to the supersymmetric 
standard model sector through gravitationally suppressed interactions.

With the sign of $A_t$ given by Eq.~(\ref{eq:sgn-At}), the constraint 
from the $b \rightarrow s \gamma$ process prefers the sign of $\mu$ 
to be positive:
\begin{equation}
  \mu > 0.
\label{eq:sgn-mu}
\end{equation}
This is because, for $A_t < 0$, the contributions from chargino and 
charged Higgs boson loops interfere destructively (constructively) 
for $\mu > 0$ ($< 0$) in the amplitude, so that the case of negative 
$\mu$ is almost excluded.  With $\mu > 0$, the constraint from the 
muon anomalous magnetic moment prefers 
\begin{equation}
  M_2 > 0.
\label{eq:sgn-M2}
\end{equation}
It is interesting that the same sign is suggested for $M_2$ as $M_3$, 
so that the simple assumption of universal gaugino masses is not 
disfavored by these considerations. 

We now consider the implication of Eqs.~(\ref{eq:mu-bound}~--~%
\ref{eq:sgn-M2}) on the detectability of neutralino dark matter. 
To naturally reproduce the correct abundance as a thermal relic, 
we assume that the lightest neutralino $\chi$ is mostly the bino, 
but containing sizable fractions of the Higgsino components:
\begin{equation}
  \chi = N_{\chi 1}\, \tilde{B} + N_{\chi 2}\, \tilde{W}^0 
       + N_{\chi 3}\, \tilde{h}_d^0 + N_{\chi 4}\, \tilde{h}_u^0,
\label{eq:chi}
\end{equation}
where $\tilde{B}$ represents the bino, $\tilde{W}^0$, $\tilde{h}_d^0$ 
and $\tilde{h}_u^0$ represent the neutral components of the wino, 
down-type Higgsino and up-type Higgsino, respectively, and the 
coefficients $N_{\chi i}$ ($i=1,\cdots,4$) are given by
\begin{eqnarray}
  N_{\chi 1} &\simeq& 1,
\\
  N_{\chi 2} &\simeq& 0,
\\
  N_{\chi 3} &\simeq& \frac{m_Z \sin\theta_W 
    (\mu \sin\beta + M_1 \cos\beta)}{\mu^2-M_1^2},
\\
  N_{\chi 4} &\simeq& -\frac{m_Z \sin\theta_W 
    (\mu \cos\beta + M_1 \sin\beta)}{\mu^2-M_1^2}.
\end{eqnarray}
Here, $m_Z$ is the $Z$ boson mass, $\theta_W$ the Weinberg angle, 
and $M_1$ the bino mass parameter.  The mass for $\chi$ is then 
given by
\begin{equation}
  m_\chi \simeq |M_1|.
\label{eq:m_chi}
\end{equation}
Note that $M_1$ can in principle take either sign, although the sign 
is positive if the three gaugino mass parameters carry the same sign, 
as in the case of universal gaugino masses.

In the parameter region relevant to us, the cross sections between 
the neutralino dark matter, $\chi$, and nuclei are dominated by the 
$t$-channel Higgs boson exchange diagrams.  There are two contributions 
coming from the lighter and heavier neutral Higgs boson exchanges, 
which are generically comparable in size.  We then find that the 
spin-independent cross section between $\chi$ and nuclei, normalized 
to the nucleon~\cite{Lewin:1995rx}, is approximately given by
\begin{equation}
  \sigma_{\rm SI} \simeq \frac{g'^4 m_N^4}{4\pi}
    \frac{\mu^2}{(\mu^2-M_1^2)^2}
    \left[ \frac{X_d \tan\beta}{m_H^2}
      + \frac{(X_u+X_d)\left(\frac{2}{\tan\beta}+\frac{M_1}{\mu}\right)}
      {m_h^2} \right]^2,
\label{eq:sigma-SI}
\end{equation}
where $g' \simeq 0.36$ is the $U(1)_Y$ gauge coupling, $m_N \simeq 
1~{\rm GeV}$ is the nucleon mass, and $X_u$ and $X_d$ are linear 
combinations of nucleon matrix elements, which we conservatively 
take as $X_u \simeq 0.14$ and $X_d \simeq 0.24$~\cite{Drees:1993bu}. 
Here, we have used the decoupling approximation for the Higgs sector, 
$\alpha \simeq \beta - \pi/2$ with $\alpha$ the neutral Higgs boson 
mixing angle, and $m_H$ and $m_h$ represent the masses of the heavier 
and lighter $CP$-even Higgs bosons, respectively.  Comparing with 
the results of more precise numerical computations, we find that the 
above simple formula reproduces the numerical results within about 
a factor of two in most of the parameter space.  These errors are 
comparable or smaller than those coming from uncertainties for the 
matrix elements.

\begin{figure}[t]
\begin{center}
  \includegraphics[height=7.5cm]{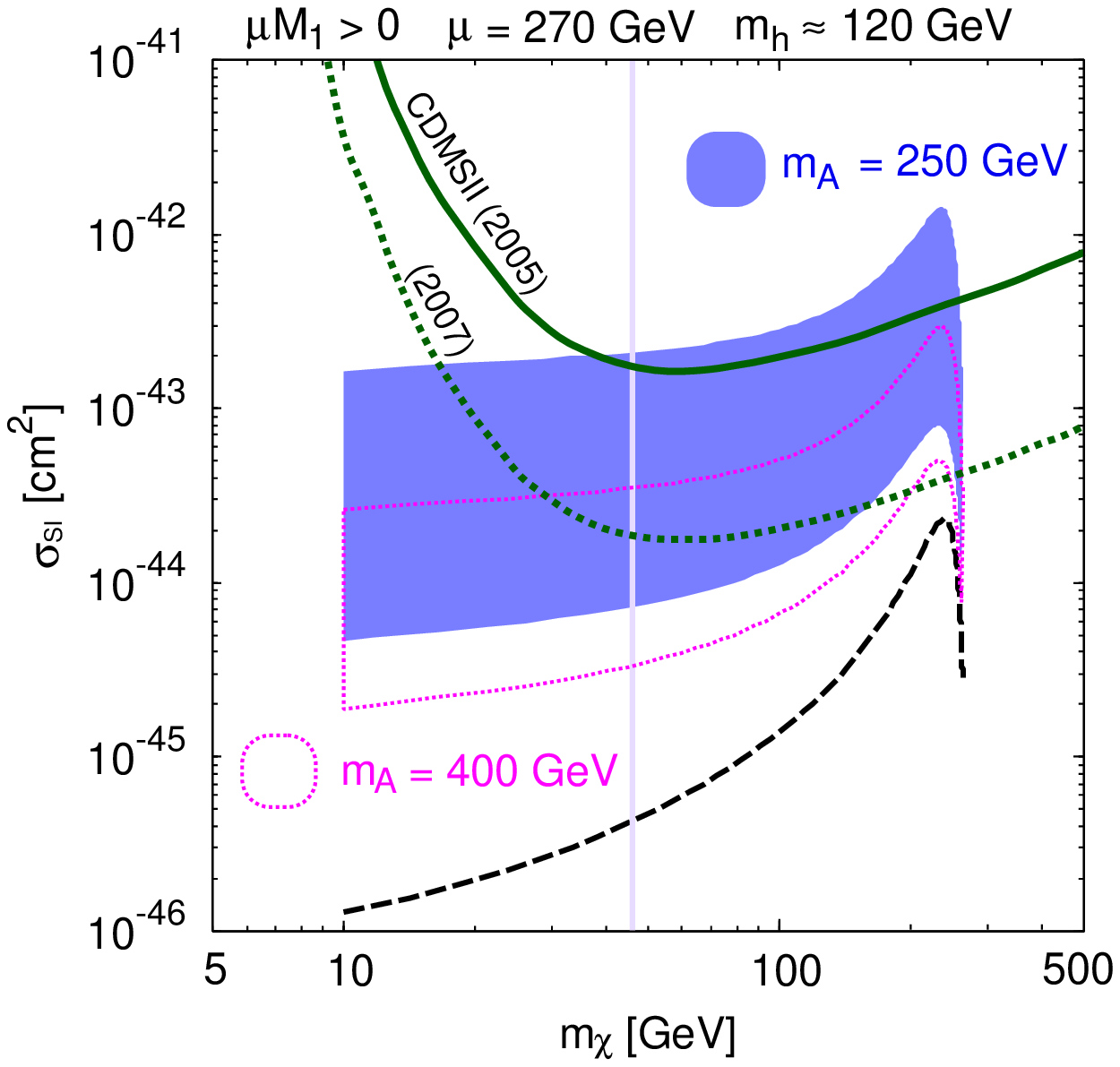}
  \includegraphics[height=7.5cm]{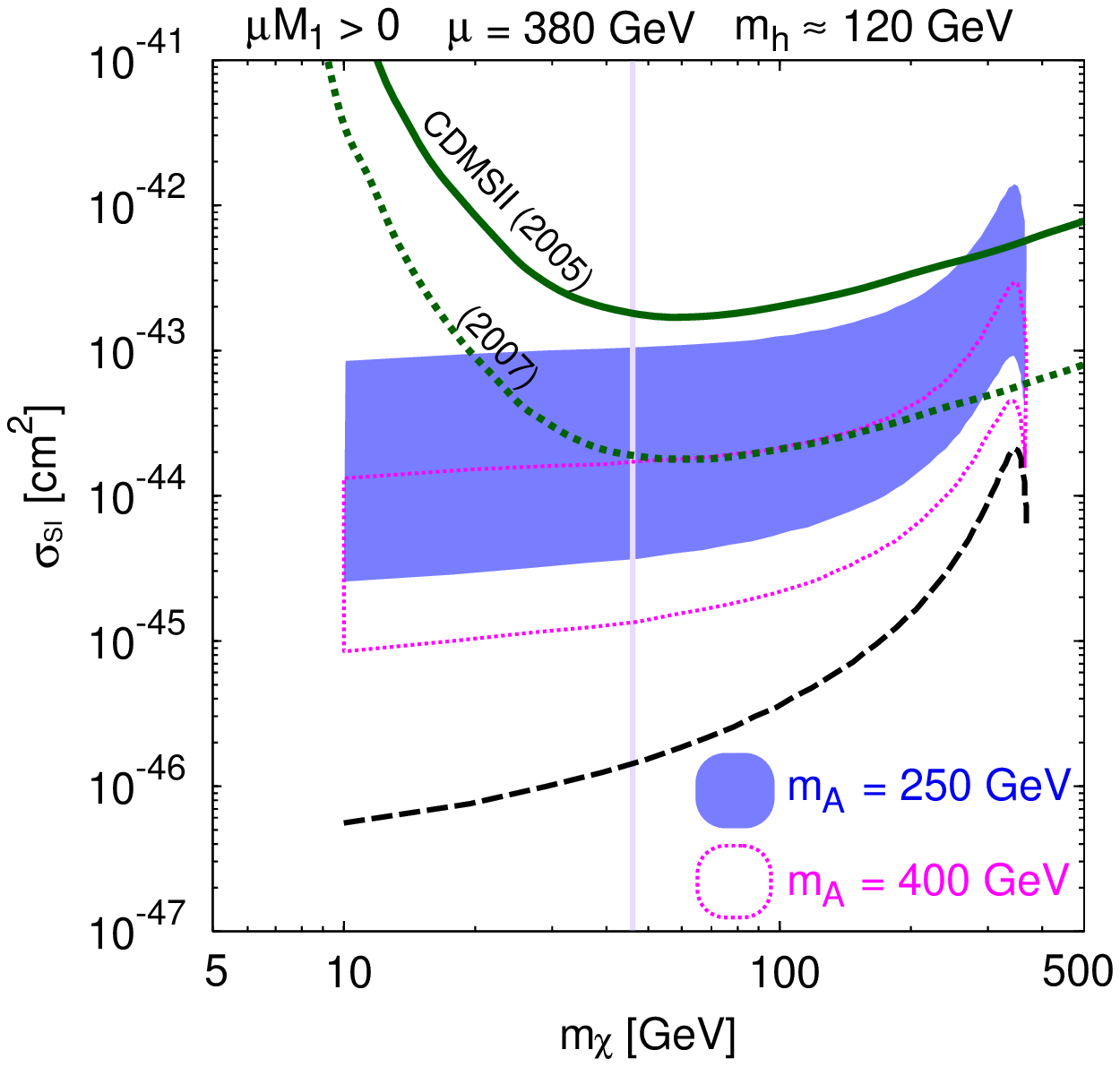}
\end{center}
\caption{The allowed range of the spin-independent cross section 
 between the dark matter and nucleon, $\sigma_{\rm SI}$, for 
 (a) $\mu = 270~{\rm GeV}$ ($\Delta^{-1} = 10\%$) and (b) $\mu = 
 380~{\rm GeV}$ ($\Delta^{-1} = 5\%$) in the case of $\mu M_1 > 0$. 
 The two regions in each plot correspond to $m_A = 250~{\rm GeV}$ 
 (shaded) and $400~{\rm GeV}$ (region inside the dotted lines), 
 and $\tan\beta$ is varied within $5 < \tan \beta < 50$.  The long 
 dashed lines correspond to the smallest possible cross section, 
 obtained for $m_A, \tan\beta \rightarrow \infty$.}
\label{fig:M1>0}
\end{figure}

The formula of Eq.~(\ref{eq:sigma-SI}) has several interesting 
implications.  First, together with Eq.~(\ref{eq:sgn-mu}), it implies 
that the contributions from the heavy and light Higgs bosons always 
interfere constructively if the sign of $M_1$ is positive, as 
in the case of universal gaugino masses.  This allows us to give 
a solid lower bound on $\sigma_{\rm SI}$ as a function of $\mu$, 
$M_1$, $\tan\beta$, $m_H$ and $m_h$.  Since $\tan\beta \simgt 5$ 
(see Eq.~(\ref{eq:tan-beta})) and $m_h \simlt 120~{\rm GeV}$, we can 
then obtain a {\it model independent} lower bound on $\sigma_{\rm SI}$ 
as a function of $m_{\chi} \simeq M_1$, once we fix the values of 
$\mu$ and $m_H$.  The value of $\mu$ is directly constrained from 
above for a given value of allowed fine-tuning $\Delta^{-1}$ (from 
Eq.~(\ref{eq:mu-bound})).  The mass of the heavy Higgs boson, $m_H$, 
does not have a definite upper bound available from the current data. 
However, since we are mainly interested in the case where all the 
superparticle masses are of order a few hundred GeV, it is natural 
to expect that $m_H$ is also not much larger than these values. 
Motivated by this, we have plotted in Fig.~\ref{fig:M1>0} the 
allowed region of $\sigma_{\rm SI}$ in the case of (a)~$\mu = 
270~{\rm GeV}$ and (b)~$\mu = 380~{\rm GeV}$, corresponding to 
$\Delta^{-1} = 10\%$ and $\Delta^{-1} = 5\%$ respectively, for 
$m_A = 250~{\rm GeV}$ and $400~{\rm GeV}$.  Here, $m_A$ is the 
mass of the pseudo-scalar Higgs boson, which is related to $m_H$ 
by $m_H^2 \simeq m_A^2 + m_Z^2 \sin^2\!2\beta$ at tree level, 
and we have used the exact formula for the Higgs-boson mediated 
dark-matter detection cross section, rather than the approximate 
formula of Eq.~(\ref{eq:sigma-SI}).
In the figure, we have also depicted the vertical line at $m_\chi = 
46~{\rm GeV}$, which is the lower bound on the mass of $\chi$ in the 
case that the gaugino masses satisfy the unified mass relations.  The 
value of $\tan\beta$ is varied between $5$ and $50$ for each value 
of $m_A$.  While the regions close to the upper edges (very large 
$\tan\beta$) are constrained by the $B_s \rightarrow \mu^+ \mu^-$ 
process~\cite{Baek:2005wi,Ellis:2006jy}, we find that it does not 
significantly affect the allowed region of $\sigma_{\rm SI}$. (The 
constraint from $b \rightarrow s \gamma$ can be satisfied depending 
on other parameters.)  In the figure, we have also drawn the 
exclusion curve from the latest CDMS~II data~\cite{Akerib:2005kh} 
by a solid line, and an estimate for the expected future sensitivity 
(an order of magnitude improvement of the current bound by the end 
of 2007~\cite{CDMS}) by a dashed line. 

\ From the figure, we find that the prospect for dark matter detection 
at CDMS~II is rather promising for $\mu = 270~{\rm GeV}$, corresponding 
to $\Delta^{-1} = 10\%$.  For $m_A = 250~{\rm GeV}$, the CDMS~II 
covers the allowed parameter region almost entirely (especially if 
$m_\chi \simgt 46~{\rm GeV}$ due to the unified gaugino mass relations), 
although we must be somewhat fortunate for larger values of $m_A$. 
For $\mu = 380~{\rm GeV}$, corresponding to $\Delta^{-1} = 5\%$, the 
discovery prospect is not as good as the case of $\mu = 270~{\rm GeV}$, 
but there is still a room that the dark matter is discovered at CDMS~II 
for $m_A \simlt 400~{\rm GeV}$.  It is interesting to point out that 
the cross section of Eq.~(\ref{eq:sigma-SI}) takes the smallest value 
at $m_H, \tan\beta \rightarrow \infty$ for fixed values of $\mu$ and 
$M_1$.  These values are depicted by long dashed lines, as a function 
of $m_\chi$, in Fig.~\ref{fig:M1>0}~(a),~(b).  We find that, for 
$\mu M_1 > 0$, the dark matter-nucleon spin-independent cross section, 
$\sigma_{\rm SI}$, due to Higgs boson exchange has an absolute lower 
bound:
\begin{equation}
  \sigma_{\rm SI} \simgt 1 \times 10^{-46}~{\rm cm}^2 
\quad (5 \times 10^{-47}~{\rm cm}^2),
\label{eq:bound-1}
\end{equation}
for $\mu \leq 270~{\rm GeV}$ ($380~{\rm GeV}$), corresponding to 
$\Delta^{-1} \simgt 10\%$ ($5\%$).  In the case that the gaugino masses 
satisfy the unified mass relations ($m_\chi \simgt 46~{\rm GeV}$), the 
bound becomes
\begin{equation}
  \sigma_{\rm SI} \simgt 4 \times 10^{-46}~{\rm cm}^2 
\quad (1 \times 10^{-46}~{\rm cm}^2).
\label{eq:bound-2}
\end{equation}
Barring the possibility of accidental cancellations with other contributions, 
such as the top squark exchange contribution, these provide {\it the 
naturalness lower bounds on the dark matter detection cross section} (for 
$\mu M_1 > 0$).  These ranges of $\sigma_{\rm SI}$ can be explored by 
future ton-scale detector experiments, such as XENON~\cite{Aprile:2002ef}. 
In fact, the allowed regions can be almost entirely covered if $m_\chi 
\simgt 46~{\rm GeV}$. 

\begin{figure}[t]
\begin{center}
  \includegraphics[height=7.5cm]{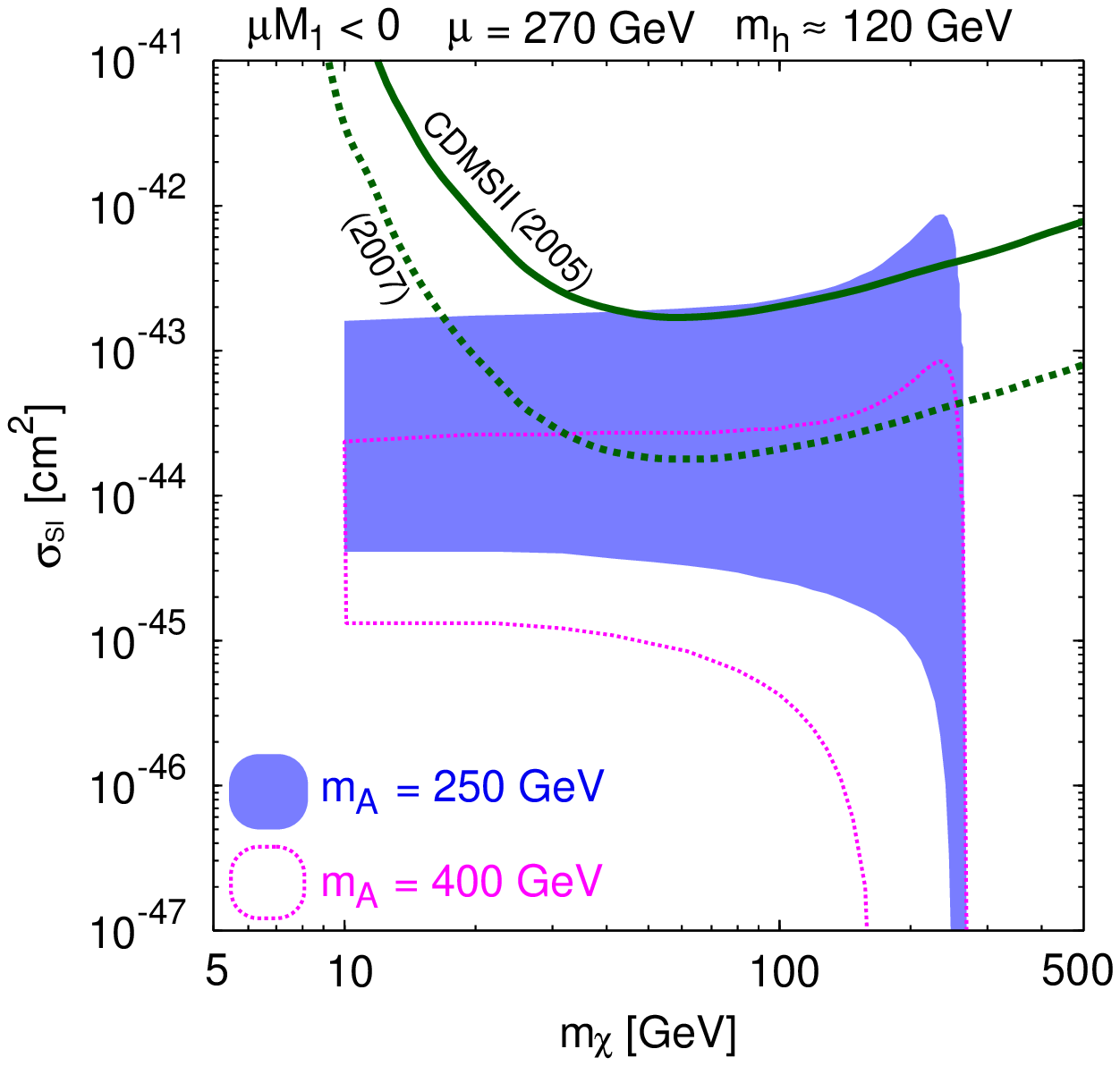}
  \includegraphics[height=7.5cm]{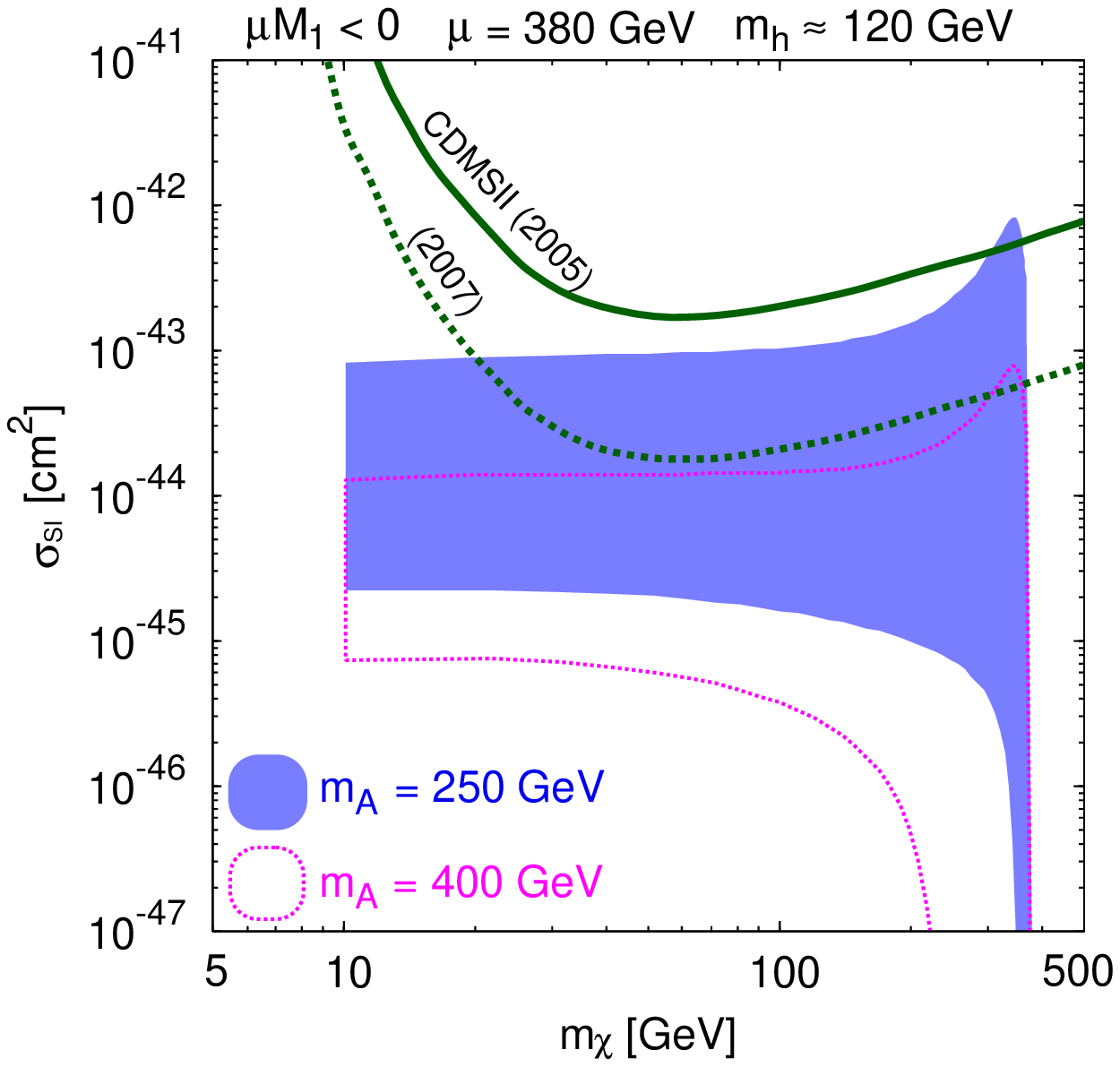}
\end{center}
\caption{The allowed range of the spin-independent cross section 
 between the dark matter and nucleon, $\sigma_{\rm SI}$, for 
 (a) $\mu = 270~{\rm GeV}$ ($\Delta^{-1} = 10\%$) and (b) $\mu = 
 380~{\rm GeV}$ ($\Delta^{-1} = 5\%$) in the case of $\mu M_1 < 0$.
 The two regions in each plot correspond to $m_A = 250~{\rm GeV}$ 
 (shaded) and $400~{\rm GeV}$ (region inside the dotted lines), 
 and $\tan\beta$ is varied within $5 < \tan \beta < 50$.}
\label{fig:M1<0}
\end{figure}
The case of $\mu M_1 < 0$, on the other hand, admits a possibility of 
cancellation between the two contributions from heavy and light Higgs 
boson exchange.  For relatively large $\tan\beta$, an excessive 
cancellation occurs for $m_h^2/m_H^2 \sim 1.5\, M_1/\mu \tan\beta$. 
In Fig.~\ref{fig:M1<0}, we have plotted the allowed region of 
$\sigma_{\rm SI}$ for $\mu M_1 < 0$ in the case of (a) $\mu = 
270~{\rm GeV}$ ($\Delta^{-1} = 10\%$) and (b) $\mu = 380~{\rm GeV}$ 
($\Delta^{-1} = 5\%$) for $m_A = 250~{\rm GeV}$ and $400~{\rm GeV}$.
Compared with the case of $\mu M_1 > 0$ in Fig.~\ref{fig:M1>0}, 
we find that the cross section can be much smaller, especially 
when the cancellation takes place, although the naturalness lower 
bound of Eq.~(\ref{eq:bound-1}) can still provide a rough guide 
on a typical range of the direct detection cross section.

Another important consequence of a small $\mu$ term is the reduction 
of the relic abundance of the lightest neutralino.  It is well known 
that the bino-dominated neutralino is over abundant unless either 
coannihilation with a slepton/stop is possible or $s$-channel diagrams 
mediated by the pseudo-scalar Higgs boson are resonantly enhanced. 
This is, however, a prediction of the ``finely tuned'' MSSM, such as 
the simplest mSUGRA, which typically gives a relatively large $\mu$ 
parameter.  Once we have a small $\mu$ term, the annihilation cross 
section into, for example, $Zh$ is significantly enhanced because it 
depends on $\mu$ by $M_1^2/\mu^4$ ($W^+W^-$ mode is proportional to 
$m_Z^4 M_1^2 / \mu^8$ for $\mu \gg M_1$).  Therefore, in models with 
{\it natural} electroweak symmetry breaking, the correct size of the 
dark matter abundance $\Omega_\chi h^2 \simeq 0.1$ can be {\it naturally} 
obtained without living in somewhat fortunate parameter regions such 
as $M_1 \simeq m_{\tilde{l}}$ or $M_1 \simeq m_A/2$.

\begin{figure}[t]
\begin{center}
  \includegraphics[height=7.5cm]{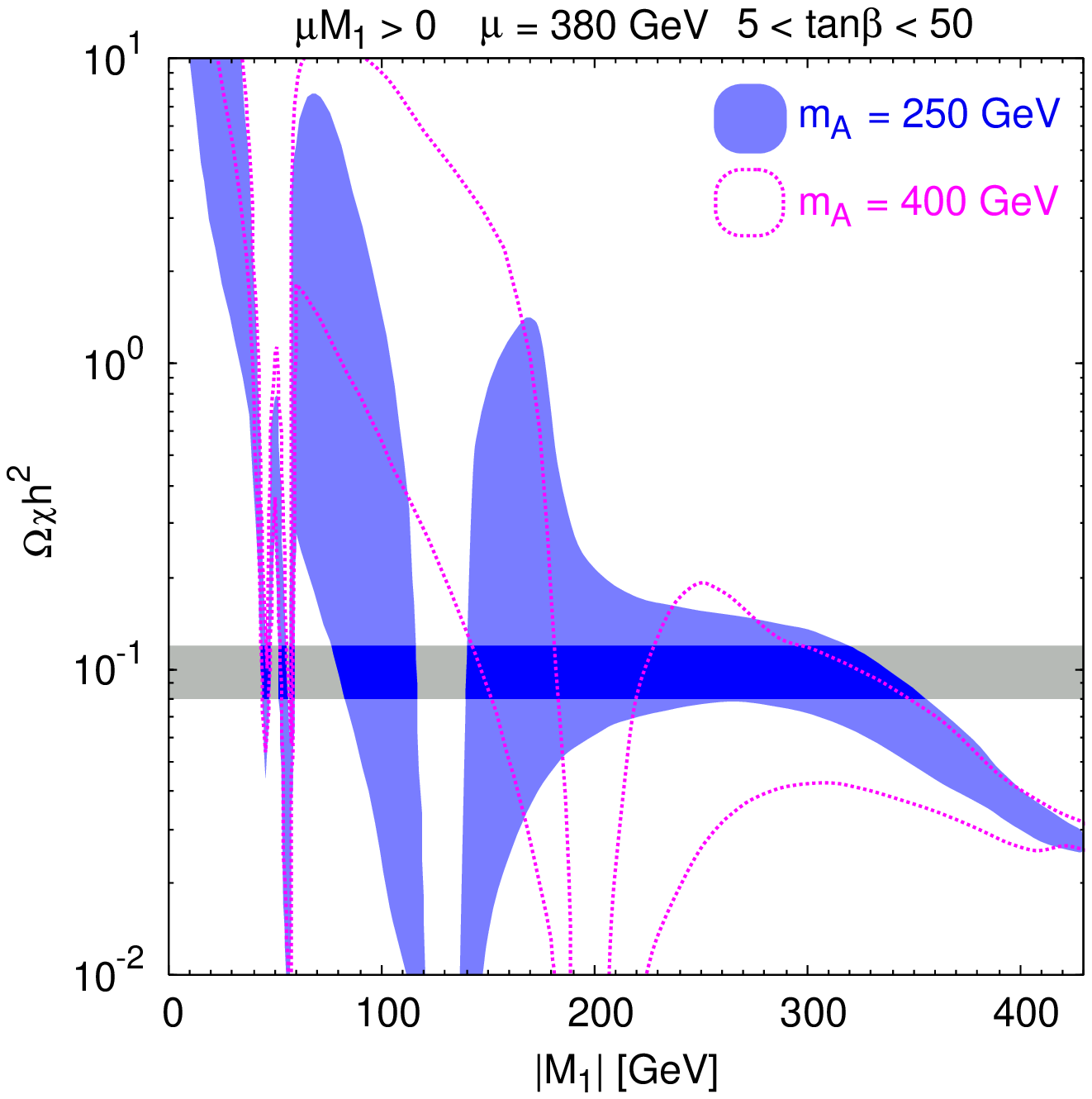}
  \includegraphics[height=7.5cm]{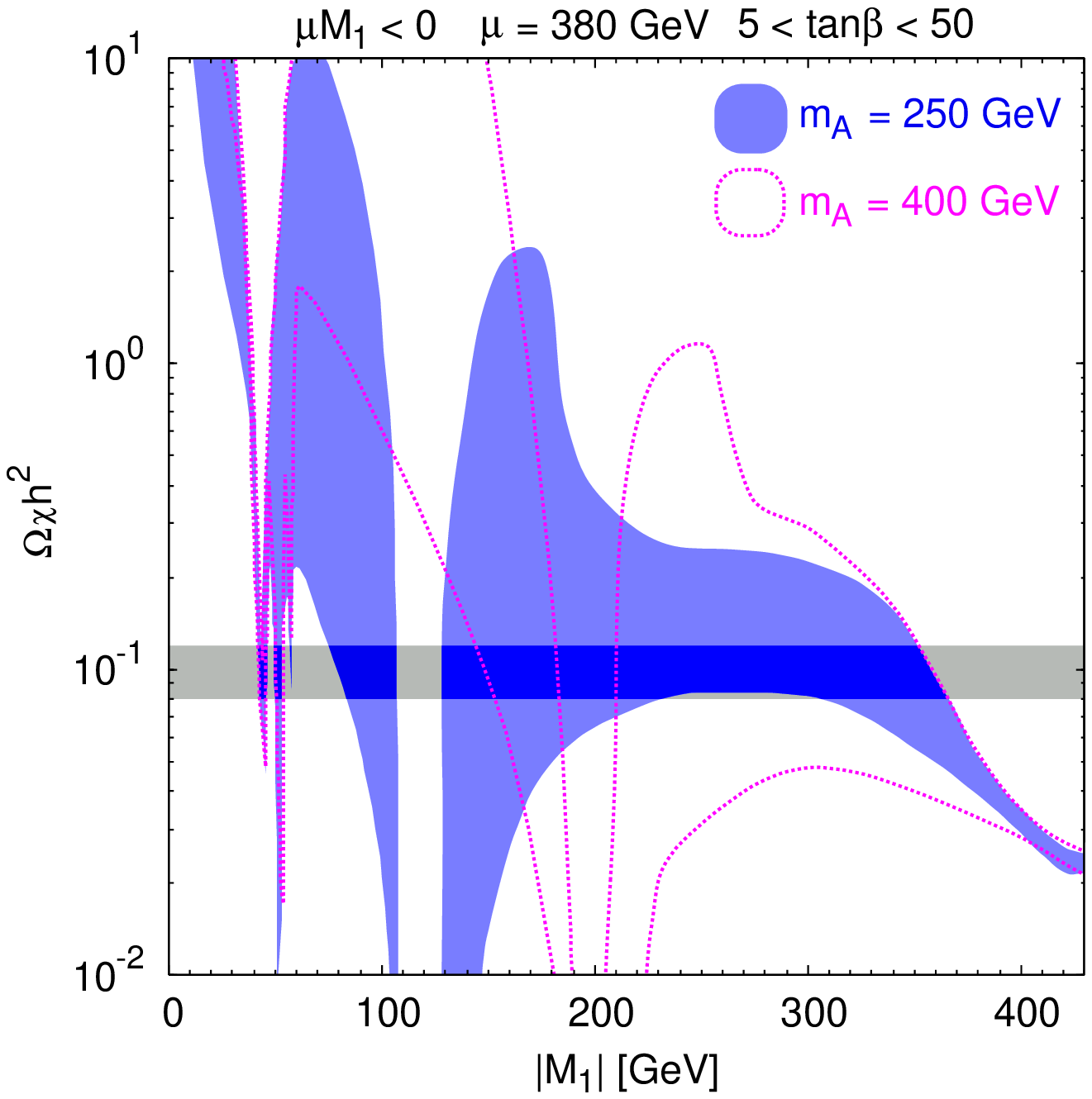}
\end{center}
\caption{The thermal relic abundance of the lightest neutralino 
 $\Omega_{\chi} h^2$ for $\mu = 380~{\rm GeV}$ for both signs of $M_1$: 
 $\mu M_1 > 0$ (left) and $\mu M_1 < 0$ (right).  The two regions in 
 each plot correspond to $m_A = 250~{\rm GeV}$ (shaded) and $400~{\rm GeV}$ 
 (region inside the dotted lines), and $\tan\beta$ is varied within 
 $5 < \tan \beta < 50$.  The WMAP data for the dark matter energy 
 density $0.08 < \Omega_{\rm DM} h^2 < 0.12$ is also indicated.}
\label{fig:M1-omega}
\end{figure}
As an example, we show in Fig.~\ref{fig:M1-omega} the thermal relic 
abundance $\Omega_\chi h^2$ of the neutralino for $\mu = 380~{\rm GeV}$ 
for both signs of $M_1$: $\mu M_1 > 0$ (left) and $\mu M_1 < 0$ 
(right).  The relic abundance has been calculated using the DarkSUSY 
package~\cite{Gondolo:2004sc}.  The value of $\tan\beta$ is varied 
within $5 < \tan \beta < 50$.  In the calculation, we have neglected 
the effect of the slepton/squark mediated annihilation processes as well 
as coannihilation effects, which depend on additional parameters.  This 
does not affect the value of $\Omega_\chi h^2$ significantly, unless 
sleptons/squarks are very light or degenerate with the lightest neutralino. 
For each value of $m_A$, we can see the effect of the $A$-pole resonance 
when $|M_1| \sim m_A/2$.  (We also see the effects of the $h$-pole 
and $Z$-pole resonances at around $|M_1| \sim 50~{\rm GeV}$.)  We find 
that values of $\Omega_\chi h^2$ consistent with the WMAP data ($0.08 
< \Omega_{\rm DM} h^2 < 0.12$ at the $2\sigma$ level) are obtained for 
$M_1$ that are not necessarily close to the pole.  In fact, we find 
that quite wide ranges of $M_1$ accommodate the observed value of 
$\Omega_\chi h^2$, due to the smallness of the $\mu$ parameter. 
(For larger values of $m_A$, the regions with the $A$-pole resonance 
disappear from the range of the figure, but we can still reproduce the 
observed value of $\Omega_{\rm DM} h^2$ naturally with $|M_1| \simlt 
400~{\rm GeV}$.  For example, for $m_A = 800~{\rm GeV}$, the WMAP 
range of $\Omega_{\rm DM} h^2$ can be reproduced for $|M_1| \approx 
(300\!\sim\!400)~{\rm GeV}$ ($\approx (300\!\sim\!350)~{\rm GeV}$) 
for $\mu M_1 > 0$ ($< 0$).)

Once we assume that the mechanism of the dark matter production is 
(dominantly) the thermal one, we can further constrain the predicted 
regions for the dark matter detection cross section, given in 
Figs.~\ref{fig:M1>0} and \ref{fig:M1<0}. 
\begin{figure}[t]
\begin{center}
  \includegraphics[height=7.5cm]{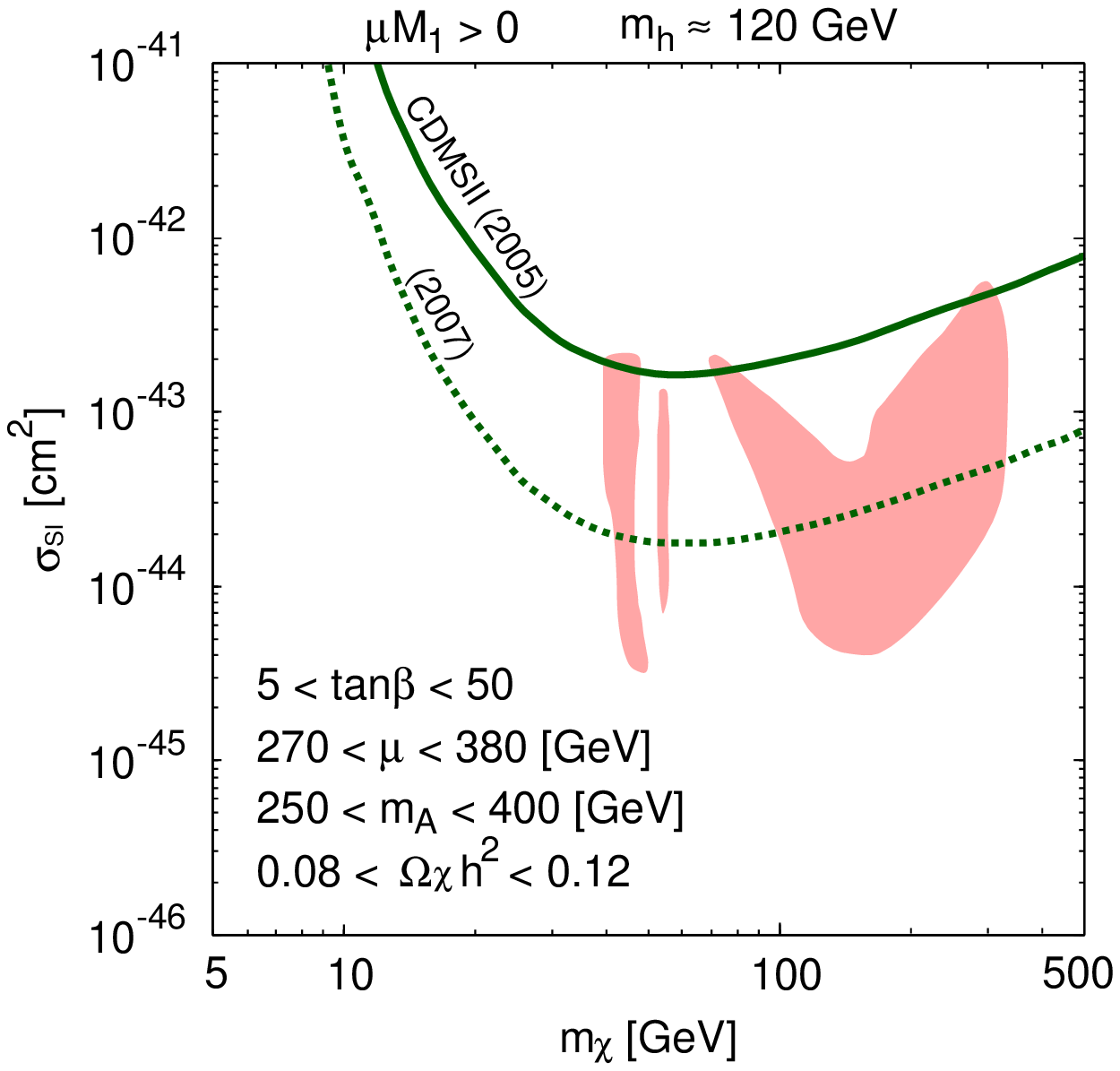}
  \includegraphics[height=7.5cm]{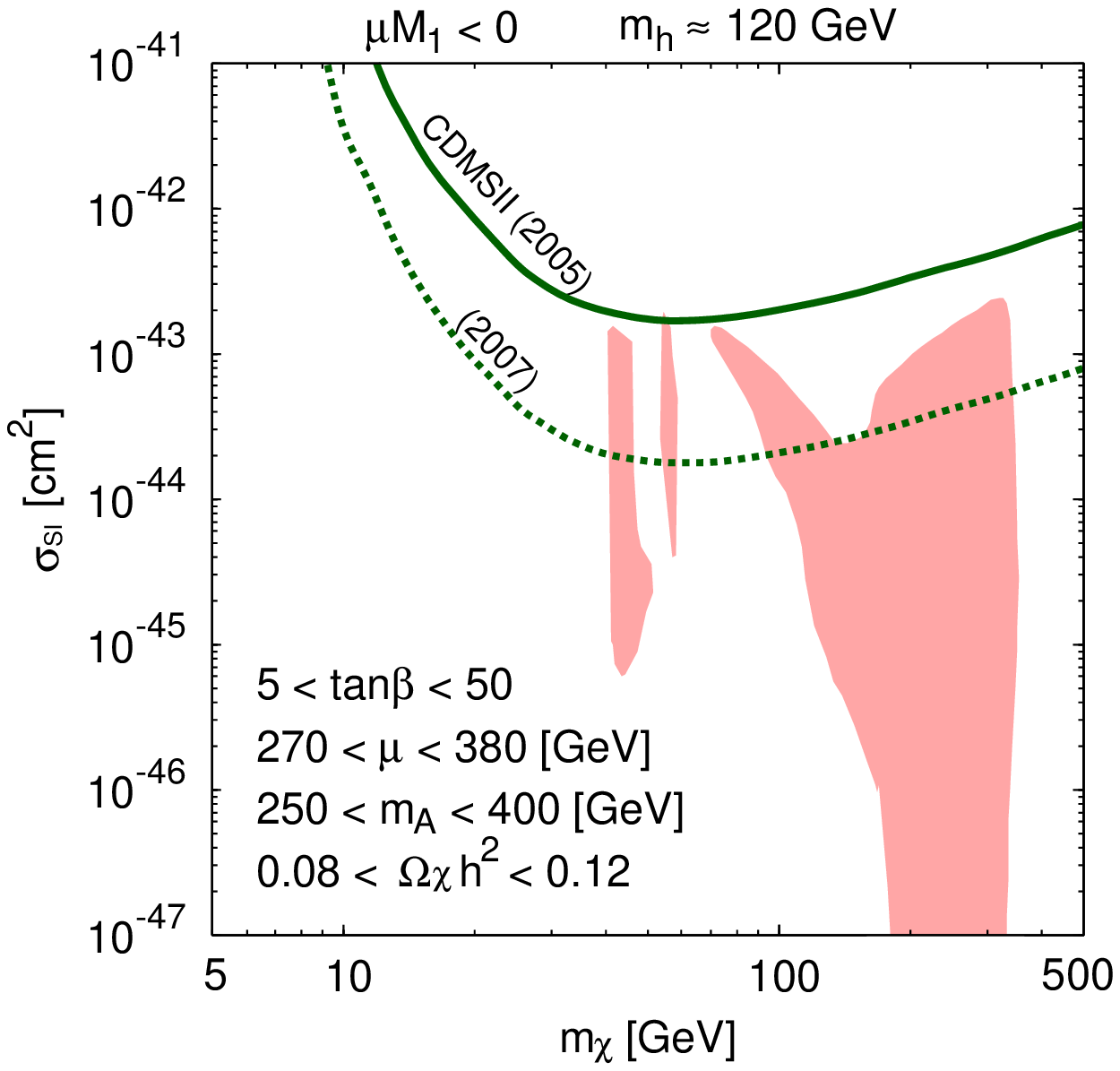}
\end{center}
\caption{The spin-independent cross section between the dark matter and 
 nucleon, $\sigma_{\rm SI}$, for $\mu M_1 > 0$ (left) and $\mu M_1 < 0$ 
 (right).  The parameters are scanned in the range $5 < \tan \beta < 50$, 
 $270~{\rm GeV} < \mu < 380~{\rm GeV}$, and $250~{\rm GeV} < m_A < 
 400~{\rm GeV}$, and only the parameter sets consistent with 
 $0.08 < \Omega_{\chi} h^2 < 0.12$ are used to draw the regions.}
\label{fig:M1-omega-sigma}
\end{figure}
In Fig.~\ref{fig:M1-omega-sigma}, we have shown the allowed regions 
in the $m_\chi$-$\sigma_{\rm SI}$ plane, under the assumption that 
the correct dark matter abundance, $0.08 < \Omega_{\chi} h^2 < 0.12$, 
is obtained thermally ($\mu M_1 > 0$ in the left panel and $\mu M_1 < 0$ 
in the right).  In drawing these regions, we have scanned $5 < \tan\beta 
< 50$, $270~{\rm GeV} < \mu < 380~{\rm GeV}$, and $250~{\rm GeV} < m_A 
< 400~{\rm GeV}$.  In each case of $\mu M_1 > 0$ and $< 0$, we can 
clearly see that only a part of the region in Fig.~\ref{fig:M1>0} 
(or Fig.~\ref{fig:M1<0}) survives.  In the case of $\mu M_1 > 0$, 
for example, the region with small $m_\chi$ disappears, and we obtain 
a lower bound of $\sigma_{\rm SI}$: $\sigma_{\rm SI} \simgt 3 \times 
10^{-45}~{\rm cm}^2$ for $m_A < 400~{\rm GeV}$.

A simple analysis presented above shows a remarkably close connection 
between naturalness of electroweak symmetry breaking and physics of 
dark matter.  This interplay can be used to narrow down parameter 
space of weak scale supersymmetry effectively.  Suppose, for example, 
that the mass and the cross section with a nucleon are measured 
for the dark matter by a direct detection experiment.  This will 
give information on the Higgs sector parameters, such as $m_A$ and 
$\tan\beta$, which are not easy to measure at the LHC.  On the other 
hand, in the parameter region under consideration, the LHC can be 
very powerful in determining the structure of the neutralino sector, 
as was shown e.g. in~\cite{Drees:2000he}.  Indirect detection of dark 
matter, for example at GLAST~\cite{Morselli:2002nw}, is also very 
promising in this parameter region.  The key point is relatively 
small values for the $\mu$ parameter.  As we have seen, this is 
required both from naturalness of electroweak symmetry breaking and 
dark matter physics.  It is interesting that this parameter region 
is also favorable for the LHC and for future dark matter detection 
experiments.

In the next section, we present a scenario that naturally leads to 
the parameter region identified in this section, in the framework of 
gravity mediation (broadly defined).  We find that the scenario has 
nontrivial implications on physics at the gravitational scale as well 
as on low-energy flavor violating processes.

\section{Next to Minimal Supergravity: Connection between Supersymmetry 
 Breaking and Flavor}
\label{sec:model}

In this section we consider possible implications of the observations 
made in the last section.  A key point for making gravity mediation 
work better is to render the low-energy value of the $\mu$ parameter 
small.  It is well known that this can be achieved by making $m_{H_u}^2$ 
larger than the other scalar squared masses at the gravitational (or 
unification) scale.  This is because the low-energy value of $m_{H_u}^2$ 
then becomes less negative, compared with the simplest mSUGRA case, 
which in turn leads to smaller values for the $\mu$ parameter (see 
Eq.~(\ref{eq:ewsb})).  Non-universal gaugino masses also help in 
this respect.  For $M_3$ smaller than $M_1$ and $M_2$ at the high 
energy threshold, the low-energy values of the top squark masses 
are smaller than in the mSUGRA case, leading to less negative 
$m_{H_u}^2$, and thus smaller $\mu$, at the weak scale. 

We note that while making $\mu$ small is a {\it necessary} condition 
to reduce fine-tuning, it is certainly not a {\it sufficient} condition. 
For instance, if we need a cancellation to make a low-energy value of 
$m_{H_u}^2$ less negative (which leads to small $\mu$), then it means 
that we simply moved the ``place'' where a cancellation/fine-tuning 
takes place.  The improvement of fine-tuning in the case of large 
$m_{H_u}^2$ at a high scale is thus nontrivial.  It arises from the 
fact that the contribution from top-stop loop is effectively reduced 
in renormalization group evolutions from high to low scales.  While 
the resulting reduction of fine-tuning is mild in this case, here 
we take it as an example of theories achieving (partially) the goals 
envisioned in the previous section and study its possible implications 
on low energy physics.%
\footnote{We emphasize, however, that the analysis in the previous 
section is much more general and applies to any theories in which 
the low energy value of $\mu$ ($> M_1$) is small.  For instance, such 
a spectrum can be obtained by making some of the (third generation) 
squark squared masses small (or even negative) at the unification 
scale, as discussed in~\cite{Kitano:2006gv,Dermisek:2006ey}.  In fact, 
this setup can be trivially accommodated in the framework discussed 
in this subsection.  Alternative possibilities include lowering 
the effective messenger scale (to some intermediate scales) by 
mixing moduli and anomaly mediated contributions to supersymmetry 
breaking~\cite{Choi:2005uz,Kitano:2006gv}.}

What could the underlying reason be, leading to $m_{H_u}^2$ larger 
than the other scalar squared masses?  An interesting possibility 
is that supersymmetry breaking and electroweak symmetry breaking 
(the Higgs fields) reside ``at the same location'' in some ``space.'' 
In this case, the observed Yukawa couplings imply that the third 
generation matter lives ``closer'' to this location, while the lighter 
generations live ``far away'' from it.  Such a setup suppresses the 
Yukawa couplings for light generations, keeping those for the third 
generation unsuppressed.  This also suppresses direct (possibly flavor 
violating) contributions in the supersymmetry breaking masses of the 
light generation sfermions.  The masses for these particles are then 
generated through standard model gauge interactions, avoiding the 
supersymmetric flavor problem.  The Higgs bosons and the third 
generation sfermions (top squarks), on the other hand, can obtain 
direct contributions from supersymmetry breaking.  Thus, they 
can have different supersymmetry breaking masses than those of 
the light generation sfermions.  This setup, therefore, reproduces 
the pattern suggested by the low energy data. 

The ``space'' described above can be real geometrical spacetime. 
For example, it may be extra space dimensions which are slightly 
larger than the fundamental scale, e.g. string scale, in which 
case direct interactions between fields localized in different 
positions are suppressed.  The resulting suppressions are 
$\approx e^{-d^2}$ ($e^{-d}$) if the interactions are generated 
by stringy effects~\cite{Hamidi:1986vh} (by exchange of massive 
fields~\cite{Randall:1998uk}), although the suppression factors are, 
in general, arbitrary~\cite{Arkani-Hamed:1999dc}.  Alternatively, 
the ``separations'' in ``space'' may be obtained effectively as 
a result of strong (nearly conformal) gauge dynamics, giving the 
Higgs and supersymmetry breaking fields as composite states.  The 
low-energy Yukawa couplings and supersymmetry breaking operators 
are then obtained through mixings between elementary and composite 
quark and lepton states.  In fact, this latter picture is obtained 
as a 4D ``dual'' picture of the former, if the extra dimension is 
one dimensional and warped~\cite{Arkani-Hamed:2000ds}.%
\footnote{A similar ``separation'' phenomenon may also be obtained 
through direct couplings of matter fields to strong conformal gauge 
dynamics~\cite{Nelson:2000sn}.} 
For earlier work on connecting structures of the Yukawa 
couplings and supersymmetry breaking parameters, see 
e.g.~[\ref{Nelson:2000sn:X}~--~\ref{Abe:2004tq:X}].

Instead of working out the detailed underlying mechanism, here we 
adopt a (useful) phenomenological parametrization of the situations 
described above.  This parameterization captures essential features 
of our general setup and provides a simple stage for phenomenological 
analyses.  Suppose we consider a 4D supergravity theory.  We assume 
that all the interactions in the superspace K\"ahler density ${\cal F}$, 
the gauge kinetic functions $f^a$, and the superpotential $W$ are of 
order unity in units of the fundamental scale $M_* \sim M_{\rm Pl}$, 
except that the (non-holomorphic) quadratic terms in ${\cal F}$ have 
arbitrary ``wavefunction factors.''  Here, $M_{\rm Pl}$ is the 4D 
reduced Planck scale, and ${\cal F}$ is related to the K\"ahler 
potential $K$ by $K = -3 M_{\rm Pl}^2 \ln(-{\cal F}/3M_{\rm Pl}^2)$. 
Denoting the supersymmetry breaking field as $X$, this leads to 
the following form for ${\cal F}$, $f^a$ and $W$:
\begin{eqnarray}
  {\cal F} &=& -3 M_{\rm Pl}^2
    + \sum_r {\cal Z}_r \Phi_r^\dagger \Phi_r 
    + {\cal Z}_X X^\dagger X 
    + \sum_r (X+X^\dagger) \Phi_r^\dagger \Phi_r
    + \sum_r X^\dagger X \Phi_r^\dagger \Phi_r + \cdots,
\label{eq:F} \\
  f^a &=& \frac{1}{g_a^2} + X + \cdots,
\label{eq:fa} \\
  W &=& Q U H_u + Q D H_d + L E H_d\,\,\, (+ L N H_u)
\nonumber\\
    && {} + X Q U H_u + X Q D H_d + X L E H_d\,\,\, (+ X L N H_u),
\label{eq:W}
\end{eqnarray}
where we have set $M_* = 1$ and omitted order one coefficients. 
The chiral superfields $\Phi_r$ represent the MSSM fields: $3 \times 
\{Q, U, D, L, E\}$, $H_u$ and $H_d$ (and $3 \times N$ if we introduce 
right-handed neutrinos).  The superscript $a$ in $f^a$ denotes 
$SU(3)_C$, $SU(2)_L$ and $U(1)_Y$, with $g_a$ the corresponding gauge 
coupling, and ${\cal Z}_r$ and ${\cal Z}_X$ are the ``wavefunction 
factors'' for the MSSM fields and the field $X$, respectively. 
The generation indices are omitted in the superpotential.

The structure given in Eqs.~(\ref{eq:F}~--~\ref{eq:W}) leads, after 
canonically normalizing fields, to the following pattern for the 
Yukawa couplings:
\begin{eqnarray}
  && (y_u)_{ij} \approx \epsilon_{Q_i} \epsilon_{U_j} \epsilon_{H_u},
\qquad
  (y_d)_{ij} \approx \epsilon_{Q_i} \epsilon_{D_j} \epsilon_{H_d},
\label{eq:yukawa-1} \\
  && (y_e)_{ij} \approx \epsilon_{L_i} \epsilon_{E_j} \epsilon_{H_d},
\qquad
  ((y_\nu)_{ij} \approx \epsilon_{L_i} \epsilon_{N_j} \epsilon_{H_u}),
\label{eq:yukawa-2}
\end{eqnarray}
where $\epsilon_r \equiv {\cal Z}_r^{-1/2}$, $i,j, = 1,2,3$ represent 
the generation indices, and $y_u$, $y_d$, $y_e$ and $y_\nu$ are 
defined by
\begin{equation}
  W = (y_u)_{ij} Q_i U_j H_u + (y_d)_{ij} Q_i D_j H_d
    + (y_e)_{ij} L_i E_j H_d\,\,\, (+ (y_\nu)_{ij} L_i N_j H_u).
\label{eq:y-def}
\end{equation}
The soft supersymmetry breaking parameters and the supersymmetric 
Higgs mass, $\mu$, take the following form.  For the non-holomorphic 
scalar squared masses, we have
\begin{equation}
  m_{\tilde{f}}^2 \approx m_0^2 + \epsilon_{f}^2 \epsilon_X^2 m^2 
    + \epsilon_{f}^4 \epsilon_X^2 m^2,
\label{eq:mf2}
\end{equation}
\begin{equation}
  m_{H_u}^2 \approx m_0^2 + \epsilon_{H_u}^2 \epsilon_X^2 m^2
    + \epsilon_{H_u}^4 \epsilon_X^2 m^2,
\qquad
  m_{H_d}^2 \approx m_0^2 + \epsilon_{H_d}^2 \epsilon_X^2 m^2
    + \epsilon_{H_d}^4 \epsilon_X^2 m^2,
\label{eq:mH2}
\end{equation}
where $f = Q_i, U_i, D_i, L_i, E_i$ (and $N_i$), and $m$ represents a 
generic mass of order $F_X/M_*$ with $F_X$ being the auxiliary field 
VEV in the canonically normalized basis.  Here, we have added a universal 
scalar squared mass term $m_0^2$.  This term does not arise from 
Eqs.~(\ref{eq:F}~--~\ref{eq:W}), but it may appear in general through 
flavor universal mediations across the ``space.''  For example, the 
gravity force is proportional to the wavefunction factors ${\cal Z}_r$ 
and ${\cal Z}_X$, so there may be a term ${\cal Z}_r {\cal Z}_X 
X^\dagger X \Phi_r^\dagger \Phi_r$ whose flavor structure is aligned 
to the kinetic terms ${\cal Z}_r$.  It may also appear through exchange 
of flavor universal bulk states in the extra dimensional setup.

The limit of a large ${\cal Z}_X$ factor (without the $m_0^2$ term) 
corresponds to the standard ``sequestering'' case, and this class 
of scenarios is widely studied as a solution to the supersymmetric 
flavor problem~\cite{Randall:1998uk,Luty:2001jh}.  Here we take 
instead a similar but different approach to solving the flavor 
problem, by sequestering light generations from the supersymmetry 
breaking sector by making ${\cal Z}_r$ large.  This setup has a virtue 
that the fermion mass hierarchy is simultaneously explained (see 
Eqs.~(\ref{eq:yukawa-1},~\ref{eq:yukawa-2})).  Given that the top 
quark has an $O(1)$ Yukawa coupling, ${\cal Z}_r$ for the (up-type) 
Higgs field should not be large, which is a desired situation from 
naturalness of the electroweak symmetry breaking. 

For the $A$ parameters, we have
\begin{eqnarray}
  (A_u)_{ij} \approx \epsilon_X m
    + (\epsilon_{Q_i}^2 + \epsilon_{U_j}^2 + \epsilon_{H_u}^2)
      \epsilon_X m,
\qquad
  (A_d)_{ij} \approx \epsilon_X m
    + (\epsilon_{Q_i}^2 + \epsilon_{D_j}^2 + \epsilon_{H_d}^2)
      \epsilon_X m,
\label{eq:A-1} \\
  (A_e)_{ij} \approx \epsilon_X m
    + (\epsilon_{L_i}^2 + \epsilon_{E_j}^2 + \epsilon_{H_d}^2)
      \epsilon_X m,
\qquad
  ((A_\nu)_{ij} \approx \epsilon_X m
    + (\epsilon_{L_i}^2 + \epsilon_{N_j}^2 + \epsilon_{H_u}^2)
      \epsilon_X m).
\label{eq:A-2}
\end{eqnarray}
Here, we have assumed that the superpotential terms containing $X$ (the 
second line in Eq.~(\ref{eq:W})) are present.  If these terms are absent 
for some reason, the first term in each expression disappears.  (Our 
definition for the $A$ parameters is such that a scalar trilinear 
coupling is given by the product of the Yukawa coupling and the $A$ 
parameter, e.g., ${\cal L} = - \sum_{i,j} (y_u)_{ij} (A_u)_{ij} \tilde{q}_i 
\tilde{u}_j H_u + {\rm h.c.}$.)  In addition, the flavor universal 
contribution, $A_0$, may be present for the same reason as $m_0^2$ in 
the scalar masses, which we have omitted in the above formulae. 

The gaugino masses are given by
\begin{equation}
  M_1 \approx \epsilon_X m,
\qquad
  M_2 \approx \epsilon_X m,
\qquad
  M_3 \approx \epsilon_X m.
\label{eq:gauginos}
\end{equation}
Note that $m$ represents a generic mass of order $F_X/M_*$, and we 
are not necessarily limiting ourselves to the case of universal gaugino 
masses.

The $\mu$ and $\mu B$ terms are generated by the terms $H_u H_d$, 
$(X+X^\dagger)H_u H_d$ and $X^\dagger X H_u H_d$ in ${\cal F}$, 
which are not explicitly denoted in Eq.~(\ref{eq:F}).  They are 
given by
\begin{equation}
  \mu \approx \epsilon_{H_u} \epsilon_{H_d} m_{3/2}
    + \epsilon_{H_u} \epsilon_{H_d} \epsilon_X m, 
\label{eq:mu}
\end{equation}
\begin{equation}
  \mu B \approx \epsilon_{H_u} \epsilon_{H_d} m_{3/2}^2
    + \epsilon_{H_u} \epsilon_{H_d} \epsilon_X m\, m_{3/2}
    + \epsilon_{H_u} \epsilon_{H_d} \epsilon_X^2 m^2,
\label{eq:muB}
\end{equation}
where $m_{3/2}$ represents a generic mass of order the gravitino 
mass, and the first, second (and third) terms in Eq.~(\ref{eq:mu}) 
(Eq.~(\ref{eq:muB})) arise from $H_u H_d$, $(X+X^\dagger)H_u H_d$ 
and $X^\dagger X H_u H_d$ in ${\cal F}$, respectively.  Here, we 
have assumed $\langle X \rangle = 0$ and a generic value (phase) 
of $F_X$, for simplicity.

The structures given in Eqs.~(\ref{eq:yukawa-1},~\ref{eq:yukawa-2},%
~\ref{eq:mf2}~--~\ref{eq:muB}) represent the results of our particular 
assumption of Eqs.~(\ref{eq:F}~--~\ref{eq:W}).  In fact, this provides 
a parameterization for very large classes of theories, larger than 
the naive picture described above (i.e. the Higgs and supersymmetry 
breaking reside in the same ``location'').  This allows us to consider 
various interesting scenarios for supersymmetry breaking.  For example, 
we can take $\epsilon_X \approx 10^{-2}$ and set $m \approx m_{3/2} 
\approx (10\!\sim\!100)~{\rm TeV}$.  In this case, anomaly 
mediation~\cite{Randall:1998uk,Giudice:1998xp} can give comparable 
contributions to the direct contributions given above.  (The standard 
problem associated with the $\mu$ and $\mu B$ terms must be solved, 
for example by replacing $\mu$ by a singlet field VEV.)  While it is 
interesting to enumerate all these possibilities, here we instead 
concentrate on the simplest case arising from the naive picture 
that supersymmetry and electroweak symmetry breaking are in the same 
``location.'' In particular, we take $\epsilon_X \sim \epsilon_{H_u} 
\sim \epsilon_{H_d} = O(1)$ and $m \sim m_{3/2}$.  We also assume 
that all the other $\epsilon$'s are smaller than $\sim 1$.  Under 
these assumptions, the Yukawa couplings take the form
\begin{eqnarray}
  (y_u)_{ij} \approx \epsilon_{Q_i} \epsilon_{U_j},
\qquad
  (y_d)_{ij} \approx \epsilon_{Q_i} \epsilon_{D_j},
\qquad
  (y_e)_{ij} \approx \epsilon_{L_i} \epsilon_{E_j},
\qquad
  ((y_\nu)_{ij} \approx \epsilon_{L_i} \epsilon_{N_j}),
\label{eq:yukawa}
\end{eqnarray}
and the soft supersymmetry breaking and $\mu$ parameters
\begin{equation}
  M_1 \approx \epsilon_X m,
\qquad
  M_2 \approx \epsilon_X m,
\qquad
  M_3 \approx \epsilon_X m,
\label{eq:M}
\end{equation}
\begin{equation}
  (m_{\tilde{f}}^2)_{ij} \approx m_0^2 + \epsilon_{f_i} \epsilon_{f_j} m^2,
\qquad
  m_{H_u}^2 \approx m_0^2 + m^2,
\qquad
  m_{H_d}^2 \approx m_0^2 + m^2,
\label{eq:m2}
\end{equation}
\begin{eqnarray}
  (A_u)_{ij} \approx 
    m + (\epsilon_{Q_i}^2 + \epsilon_{U_j}^2 + \delta_{ij} ) m,
\qquad
  (A_d)_{ij} \approx 
    m + (\epsilon_{Q_i}^2 + \epsilon_{D_j}^2 + \delta_{ij} ) m,
\label{eq:Au-Ad} \\
  (A_e)_{ij} \approx 
    m + (\epsilon_{L_i}^2 + \epsilon_{E_j}^2 + \delta_{ij} ) m,
\qquad
  ((A_\nu)_{ij} \approx 
    m + (\epsilon_{L_i}^2 + \epsilon_{N_j}^2 + \delta_{ij} ) m),
\label{eq:Ae-Anu}
\end{eqnarray}
\begin{equation}
  \mu \approx m, 
\qquad
  \mu B \approx m^2,
\label{eq:mu-muB}
\end{equation}
where $f = Q_i, U_i, D_i, L_i, E_i$ (and $N_i$), and $m$ represents 
a generic mass parameter of order the weak scale.  Note that the first 
terms in the $A$-term formulae (Eqs.~(\ref{eq:Au-Ad},~\ref{eq:Ae-Anu})) 
arises from the superpotential terms containing $X$ (the second line 
in Eq.~(\ref{eq:W})), which can have nontrivial flavor dependences. 
The flavor universal contributions $\delta_{ij} m$ in the $A$ terms 
originate from the $\epsilon_{H_u}^2 \epsilon_X m$ or $\epsilon_{H_d}^2 
\epsilon_X m$ term, as well as from the $A_0$ term.  These equations 
determine the correlations between the structures of the Yukawa 
couplings and the soft supersymmetry breaking parameters.

Let us now study consequences of Eqs.~(\ref{eq:yukawa}~--~\ref{eq:mu-muB}). 
We first consider the case where the $SU(5)$ relations are satisfied 
at the unification scale $M_U \approx 10^{16}~{\rm GeV}$:
\begin{equation}
  \epsilon_{Q_i} = \epsilon_{U_i} = \epsilon_{E_i},
\qquad
  \epsilon_{D_i} = \epsilon_{L_i},
\label{eq:SU5-1}
\end{equation}
\begin{equation}
  M_1 = M_2 = M_3,
\label{eq:SU5-2}
\end{equation}
\begin{equation}
  m_{\tilde{Q}_i}^2 = m_{\tilde{U}_i}^2 = m_{\tilde{E}_i}^2,
\qquad
  m_{\tilde{D}_i}^2 = m_{\tilde{L}_i}^2,
\label{eq:SU5-3}
\end{equation}
\begin{equation}
  (A_d)_{ij} = (A_e)_{ij},
\qquad
  ((A_u)_{ij} = (A_\nu)_{ij}),
\label{eq:SU5-4}
\end{equation}
although this need not be the case.  (Unwanted fermion mass relations 
for the first two generations must be corrected somehow.)

With this assumption, we can determine the order of magnitude for the 
$\epsilon$ parameters from the fermion masses.  From the Yukawa coupling 
of the up-type quarks, $\epsilon^{(10)}_i$ ($\equiv \epsilon_{Q_i} = 
\epsilon_{U_i} = \epsilon_{E_i}$) are obtained as
\begin{equation}
  \epsilon^{(10)}_i
  \simeq \sqrt{(y_u)_{ii}} 
  \simeq (3 \times 10^{-3},\,\, 4 \times 10^{-2},\,\, 8 \times 10^{-1}).
\label{eq:eps-10}
\end{equation}
The $\epsilon^{(\bar{5})}_i$ ($\equiv \epsilon_{D_i} = \epsilon_{L_i}$) 
factors can be estimated by two ways; from down-type quarks or charged 
leptons.  Those are given by
\begin{equation}
  \epsilon^{(\bar{5})}_i 
  \simeq \frac{(y_d)_{ii}}{\sqrt{(y_u)_{ii}}} 
  \simeq \tan\beta 
    \times (4 \times 10^{-3},\,\, 5 \times 10^{-3},\,\, 9 \times 10^{-3}),
\label{eq:eps-5*-down}
\end{equation}
and
\begin{equation}
  \epsilon^{(\bar{5})}_i 
  \simeq \frac{(y_e)_{ii}}{\sqrt{(y_u)_{ii}}} 
  \simeq \tan\beta 
    \times (1 \times 10^{-3},\,\, 1 \times 10^{-2},\,\, 1 \times 10^{-2}).
\label{eq:eps-5*-lepton}
\end{equation}
It is interesting that the above two are very similar.  Moreover, the 
less hierarchical structure of $\epsilon^{(\bar{5})}_i$ is also consistent 
with large mixing angles and small mass hierarchies in the neutrino 
sector, provided that the neutrino masses are of the Majorana type 
$(L H_u)^2$~\cite{Yanagida:1998jk}.  Equations~(\ref{eq:eps-10}~--~%
\ref{eq:eps-5*-lepton}) imply that, unless $\tan\beta$ is very large, 
only the third generation ${\bf 10}$ multiplet, $Q_3$, $U_3$ and $E_3$, 
has an $O(1)$ $\epsilon$ factor.

\subsection{Impact on electroweak symmetry breaking}
\label{subsec:ewsb}

Supersymmetry breaking parameters in our setup are essentially 
a modification of the mSUGRA ones in the Higgs and third generation 
sfermion sectors, due to $O(1)$ $\epsilon$ factors.  Because the 
coefficients of the operators in Eqs.~(\ref{eq:F}~--~\ref{eq:W}) are 
free parameters, the relevant parameters for electroweak symmetry 
breaking, $m_{H_u}^2$, $m_{H_d}^2$, $\mu$, $\mu B$, $m_{\tilde{Q}_{3}}^2$, 
$m_{\tilde{U}_{3}}^2$, and $M_3$ at the unification scale, $M_U$, can 
all be taken as independent free parameters.  Equivalently, we can 
treat $m_A$, $\mu$, $\tan\beta$, and $v$ $(\equiv (\langle H_u \rangle^2 
+ \langle H_d \rangle^2)^{1/2} = 174~{\rm GeV})$ at a low energy as 
input parameters, instead of $m_{H_u}^2$, $m_{H_d}^2$, $\mu$, and 
$\mu B$.  The free parameters of our electroweak symmetry breaking 
analysis can thus be taken as $m_A$, $\mu$ and $\tan\beta$ at the 
weak scale, as well as $m_{\tilde{Q}_{3}}^2$, $m_{\tilde{U}_{3}}^2$ 
and $M_3$ at $M_U$.

It was shown in Ref.~\cite{Kitano:2006gv} that fine-tuning in electroweak 
symmetry breaking can be improved from the mSUGRA case by relaxing (one 
of) the relations among those parameters.  The least fine-tuned region 
requires a large $A_t$ parameter to avoid the constraint from the Higgs 
boson mass bound.  A large $A_t$ term is naturally obtained in our 
setup, even if the contribution from the superpotential term $XQUH_u$ 
and the universal contribution $A_0$ are absent, due to the terms from 
the K\"ahler potential.  (In fact, the absence of the superpotential 
term is somewhat favored by the constraints from flavor violating 
processes, as we discuss later.)  With a large $A_t$ term, the gluino 
and stop masses, $M_{\tilde{g}}$, $m_{\tilde{t}_1}$ and $m_{\tilde{t}_2}$, 
can be as low as $\approx 440~{\rm GeV}$, $\approx 120~{\rm GeV}$ and 
$\approx 430~{\rm GeV}$, respectively, without contradicting with the 
experimental constraints, including the Higgs boson mass bound.  With 
these relatively small supersymmetry breaking parameters, significant 
fine-tuning among fundamental parameters is not needed to reproduce 
the correct scale for electroweak symmetry breaking.

\begin{figure}[t]
\begin{center}
  \includegraphics[width=7.5cm]{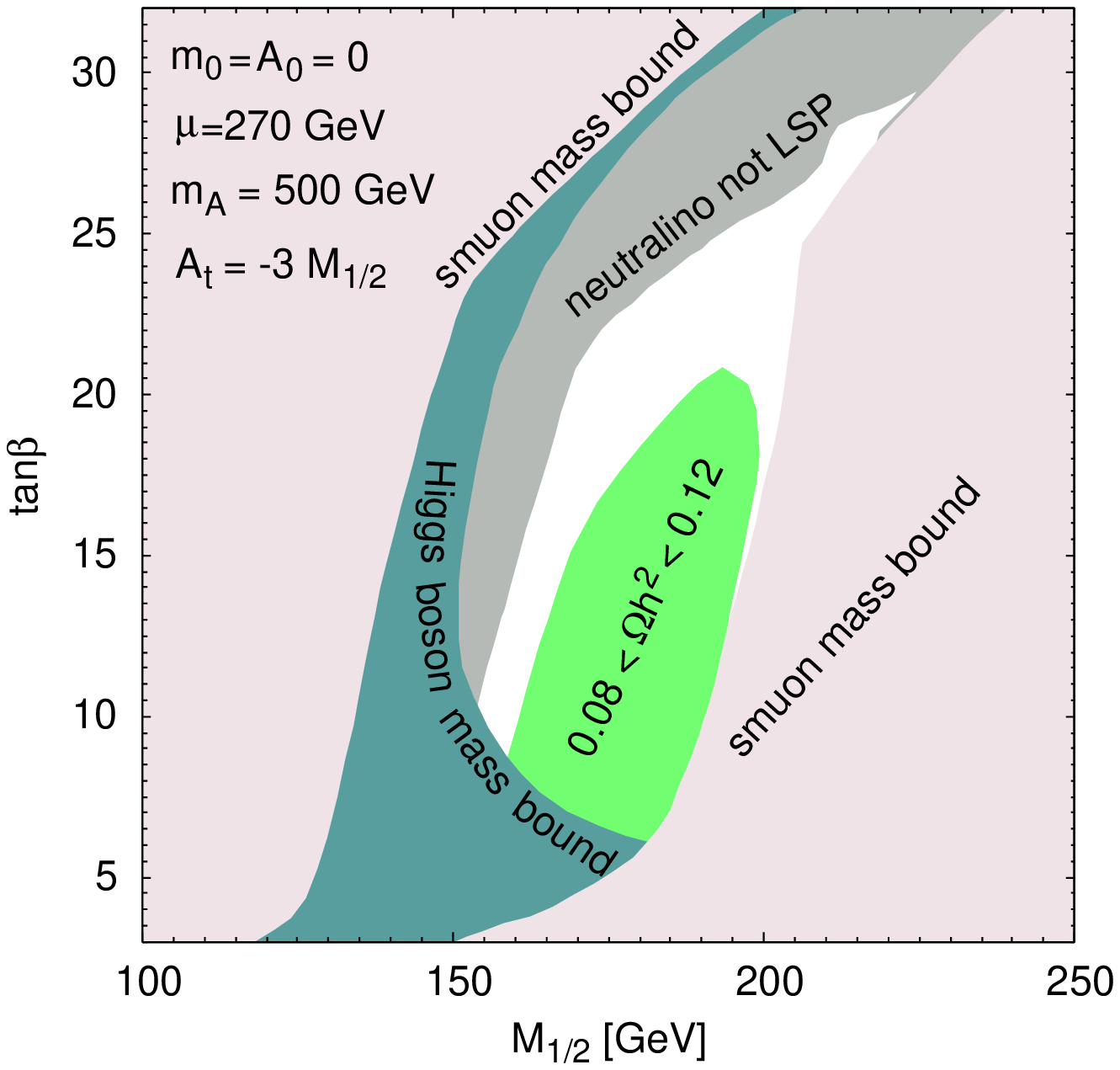}
  \includegraphics[width=7.5cm]{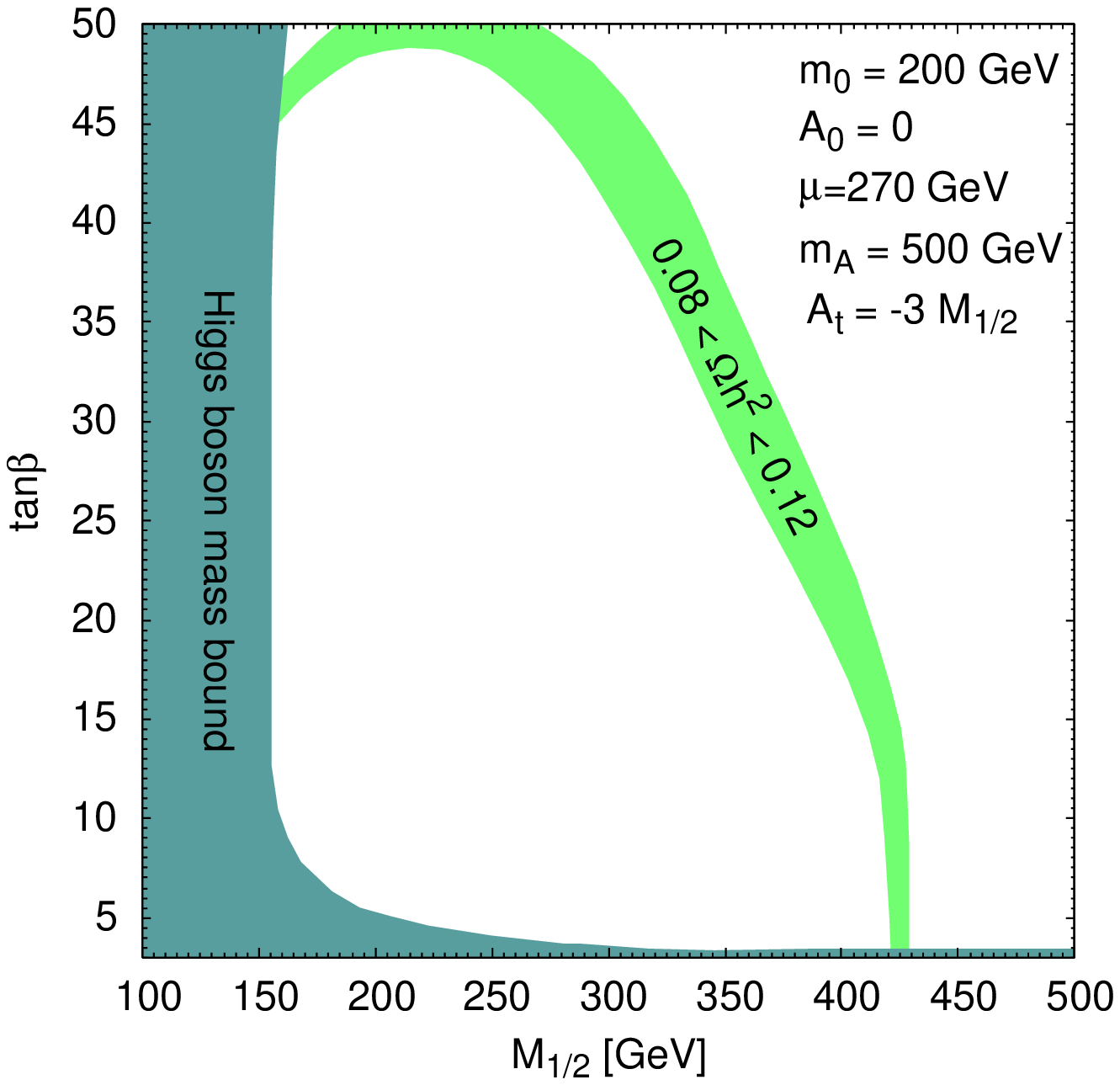}
\end{center}
\caption{Viable parameter regions for the model with $m_0 = 0$ (left) and 
 $200$~GeV (right). We have fixed the values of $\mu = 270~{\rm GeV}$ and 
 $m_A = 500~{\rm GeV}$ at low energies.  Gaugino masses at the unification 
 scale $M_{1/2}$ are taken to be universal and we take $A_t = -3 M_{1/2}$ 
 for the top squarks at the unification scale.  The regions with correct 
 dark matter abundance are also indicated.}
\label{fig:region}
\end{figure}
In Fig.~\ref{fig:region}, we show viable parameter regions for the 
cases of $m_0 = 0$ and $200~{\rm GeV}$.  The gaugino masses are taken 
to be universal $M_{1/2} \equiv M_1 = M_2 = M_3$ at $M_U$, and we 
have fixed the sign of $\mu$ to be $\mu M_{1/2} > 0$, motivated by the 
constraints from $b \to s \gamma$ and the muon anomalous magnetic moment. 
The values of third generation sfermion squared masses are chosen to 
be $m_{\tilde{Q}_{3}}^2 = m_{\tilde{U}_{3}}^2 = m_{\tilde{E}_{3}}^2 
= (\epsilon^{(10)}_3 M_{1/2})^2$ and $m_{\tilde{D}_{3}}^2 = 
m_{\tilde{L}_{3}}^2 = (\epsilon^{(10)}_3 \epsilon^{(\bar{5})}_3) 
M_{1/2}^2$ at $M_U$. 

We find that there is a parameter region for $m_0 = 0$, which may 
be interesting from a theoretical point of view, since it arises from 
a simple form of Eqs.~(\ref{eq:F}~--~\ref{eq:W}).  In fact, the existence 
of the region is nontrivial.  On one hand, $M_{1/2}$ should be larger 
than $(120\!\sim\!200)~{\rm GeV}$, since the first two generation 
sleptons obtain their masses only from one-loop running effects.  On 
the other hand, the one-loop induced $D$-term for $U(1)_Y$ gives negative 
contributions to the right-handed slepton masses if $m_{H_d}^2$ is smaller 
than $m_{H_u}^2$ at $M_U$.  With fixed values of $\mu$ and $m_A$, this 
happens when the gaugino mass is large, putting an upper bound on 
the gaugino mass.  The Higgs boson mass bound is not severe in this 
model, since we can have a large $A_t$ term, which is fixed to be 
$A_t = -3 M_{1/2}$ at $M_U$ in the figure.  The excluded region is 
indicated, where we have imposed a conservative bound of $M_{\rm Higgs} 
> 113~{\rm GeV}$ to take into account theoretical uncertainties in 
the Higgs boson mass calculation.  A relatively large value of $m_A 
= 500~{\rm GeV}$ also helps to obtain large values of $\tan\beta 
\approx (10\!\sim\!20)$ naturally.  It is quite interesting that the 
region actually exists at a relatively low supersymmetry breaking scale, 
which is desired from naturalness.  Solving the electroweak VEV, $v$, 
as a function of high energy parameter, and finding the severest 
cancellation among different contributions, we find that fine-tuning 
of the viable parameter region in Fig.~\ref{fig:region}~(left) 
is or order $(5\!\sim\!10)\%$, which is better than the simplest 
mSUGRA case.  We note that a nonvanishing (positive) value of 
$m_{\tilde{Q}_{3}}^2 = m_{\tilde{U}_{3}}^2 = m_{\tilde{E}_{3}}^2$ 
at $M_U$ helps to have values of $M_{1/2}$ as small as $\simeq 
170~{\rm GeV}$, by providing a positive contribution to the right-handed 
stau mass.  Without them, the value of $M_{1/2}$ is pushed up to above 
$\simeq 200~{\rm GeV}$ making fine-tuning somewhat worse, although 
we can still obtain a consistent parameter region in this case.%
\footnote{In fact, we do not need the entire scaling of soft supersymmetry 
breaking parameters in Eqs.~(\ref{eq:M}~--~\ref{eq:mu-muB}) to obtain 
a desired parameter region in Fig.~\ref{fig:region}~(left).  All we 
need are nonvanishing squared masses for the third generation ${\bf 10}$ 
scalars at $M_U$ (in addition to the gaugino masses and nonvanishing 
$A$ terms).  This arises in any setup where the first two generation 
(and the third generation in ${\bf 5}^*$) fields are separated from 
supersymmetry breaking while the third generation (in ${\bf 10}$) 
and Higgs fields are not.  A detailed analysis on this and related 
possibilities will be presented elsewhere.}

For $m_0 = 200~{\rm GeV}$, there is no constraint from the slepton masses. 
The bound from the Higgs boson mass is $M_{1/2} \simgt 150~{\rm GeV}$ 
for large $A_t$.  Reproducing the dark matter abundance, however, 
requires somewhat larger values of $M_{1/2}$ or $\tan\beta$, so 
electroweak fine-tuning is not as good as the case of $m_0 = 0$. 

As explained before, a small $\mu$ term is an inevitable consequence 
of naturalness in electroweak symmetry breaking, which also provides 
a natural way of explaining the observed dark matter abundance.  The 
regions with a correct dark matter abundance are superimposed in 
Fig.~\ref{fig:region}.  It is clear that we do not have to live in 
special regions of the parameter space, such as $m_\chi \sim m_A/2$. 
Moreover, since the sign of $\mu M_1$ is positive, the direct detection 
of the neutralino dark matter is promising in this scenario.

\subsection{Impact on flavor physics}
\label{subsec:flavor}

With the flavor structure of supersymmetry breaking terms deduced 
from the fermion masses through Eqs.~(\ref{eq:M}~--~\ref{eq:mu-muB}), 
we can make predictions on the magnitude of flavor violation in 
the low energy observables.  Although we cannot precisely calculate 
the rates of the processes due to $O(1)$ ambiguities, simple relations 
between the rates and the Yukawa structure can be obtained.  For 
the analysis of flavor violating processes, we follow the method 
of Ref.~\cite{Gabbiani:1996hi}, where various constraints on 
flavor mixing parameters are listed.

The most important source of flavor violation comes from off-diagonal 
components of the $A$ terms that arise from the couplings between the 
supersymmetry breaking field $X$ and the MSSM fields in the superpotential 
(the second line in Eq.~(\ref{eq:W})).  This is because, although the 
flavor mixings are suppressed by $\epsilon$ factors, these suppressions 
are compensated by the fact that an $A$-term insertion flips the 
chirality of the sfermion, and thus eliminates one factor of the 
Yukawa coupling from the amplitude.  We thus first consider the effects 
of these terms, and later consider the case where these terms are 
somehow absent.

For $\mu \to e$ transition processes, the ratio of the off-diagonal to 
the diagonal components of the mass matrix, $(\delta^l_{12})_{LR}$ and 
$(\delta^l_{12})_{RL}$, are given by
\begin{equation}
  (\delta^l_{12})_{LR} \simeq \epsilon_1^{(\bar{5})} \epsilon_2^{(10)} 
    \left( \frac{\langle H_d \rangle}{m_{\rm SUSY}} \right)
  \simeq 4 \times 10^{-5}
    \left( \frac{v}{m_{\rm SUSY}} \right),
\end{equation}
\begin{equation}
  (\delta^l_{12})_{RL} \simeq \epsilon_2^{(\bar{5})} \epsilon_1^{(10)} 
    \left( \frac{\langle H_d \rangle}{m_{\rm SUSY}} \right)
  \simeq 3 \times 10^{-5}
    \left( \frac{v}{m_{\rm SUSY}} \right),
\end{equation}
where we have used $\epsilon_1^{(\bar{5})}$ obtained from the charged 
lepton masses in Eq.~(\ref{eq:eps-5*-lepton}), and $v$ and $m_{\rm SUSY}$ 
represent the Higgs VEV, $v \simeq 174~{\rm GeV}$, and the mass scale 
of supersymmetry breaking parameters, respectively.  Note that these mass 
insertion parameters do not depend on $\tan\beta$, in contrast to many 
supersymmetric models where the $\mu \to e$ transition amplitude is 
proportional to $\tan\beta$.  Since the upper bounds on these variables 
from the branching ratio of the $\mu \to e \gamma$ decay are of order 
$10^{-6}$ (corresponding to $B(\mu \to e\gamma) \simlt 10^{-11}$), we 
need somewhat small coefficients for the superpotential couplings between 
$X$ and the MSSM fields, such as $O(0.1)$.  Large supersymmetry breaking 
masses would help to suppress the amplitude, but it is disfavored from 
naturalness in electroweak symmetry breaking.  This may imply that 
the superpotential couplings between $X$ and the MSSM fields (at least 
for light generations) are somehow absent, although small couplings 
of $O(0.1)$ may still be regarded as ``acceptable.''  In any case, if 
the superpotential couplings are present, we expect positive signals 
in future searches of $\mu \to e \gamma$ decay and the $\mu \to e$ 
conversion process in nuclei, which have sensitivities to the level 
of $O(10^{-8}\!\sim\!10^{-7})$ in these parameters~\cite{MEG,PRISM}. 

For flavor violating $\tau$ decays, the following simple relations 
can be obtained:
\begin{equation}
  \frac{B(\tau \to e \gamma)}{B(\mu \to e \gamma)} 
  \sim 0.2 \left[ \frac{\epsilon_1^{(\bar{5})} \epsilon_3^{({10})}}
    {\epsilon_1^{(\bar{5})} \epsilon_2^{({10})}} 
    \frac{m_\mu}{m_\tau} \right]^2
  \sim 0.2,
\quad
  \frac{B(\tau \to \mu \gamma)}{B(\mu \to e \gamma)} 
  \sim 0.2 \left[ \frac{\epsilon_2^{(\bar{5})} \epsilon_3^{({10})}}
    {\epsilon_1^{(\bar{5})} \epsilon_2^{({10})}}
    \frac{m_\mu}{m_\tau} \right]^2
  \sim 20.
\label{eq:mu-tau}
\end{equation}
Comparing with the current experimental sensitivities to the branching 
ratios of $O(10^{-7})$ for $\tau$ decays~\cite{Abe:2003sx,Aubert:2005ye} 
and $O(10^{-11})$ for $\mu \to e \gamma$ decay, we conclude that flavor 
violating $\tau$ decays are not likely to be observed in near future 
in this model.

Similar analyses can be performed for hadronic processes.  For the 
$K^0$-$\bar{K}^0$ mixing, the contribution from the $A$ term is simply
\begin{equation}
  (\delta_{12}^d)_{LR} \simeq  (\delta_{12}^l)_{RL},
\quad
  (\delta_{12}^d)_{RL} \simeq  (\delta_{12}^l)_{LR},
\end{equation}
from the similarity of the down-type and charged-lepton Yukawa couplings. 
We find that the constraint from $\Delta m_K$, of $O(10^{-3})$, is much 
weaker than that from $\mu \to e \gamma$ decay.  Other meson mixings 
such as $\Delta m_{B_d}$, $\Delta m_{B_s}$ and $\Delta m_{D}$ are also 
predicted to be much smaller than the experimental constraints.  For the 
gluino mediated $b \to s \gamma$ decay, the most important mass insertion 
factor is:
\begin{equation}
  (\delta^d_{23})_{RL} 
  \simeq \epsilon_2^{(\bar{5})} \epsilon_3^{(10)} 
    \left( \frac{ \langle H_d \rangle}{m_{\rm SUSY}} \right)
  \simeq 8 \times 10^{-3} \left( \frac{v}{m_{\rm SUSY}} \right).
\end{equation}
The experimental constraint of $O(10^{-2})$ is marginally satisfied.

Let us now consider the case where the superpotential couplings between 
$X$ and the MSSM fields are somehow suppressed.  This setup is technically 
natural and does not require fine-tuning between parameters.  In this 
case, contributions from the scalar mass terms in Eq.~(\ref{eq:m2}) become 
important sources of flavor violation, and we can repeat the same analysis 
as before to see the predictions.  For $\mu \to e \gamma$ decay and the 
$\mu$--$e$ conversion in nuclei, the most significant bound comes from 
diagrams with $\tan\beta$ enhanced chirality flipping.  The predictions 
of the model from such diagrams are parametrized as
\begin{equation}
  (\delta^l_{12})_{RL}^{\rm eff} 
  \simeq \epsilon_1^{(10)} \epsilon_2^{(10)} 
    \frac{m_\mu \tan \beta }{m_{\rm SUSY}}
  \simeq 4 \times 10^{-8} \tan\beta 
    \left( \frac{v}{m_{\rm SUSY}} \right),
\end{equation}
\begin{equation}
  (\delta^l_{12})_{LR}^{\rm eff} 
  \simeq \epsilon_1^{(\bar{5})} \epsilon_2^{(\bar{5})} 
    \frac{m_\mu \tan \beta }{m_{\rm SUSY}}
  \simeq 5 \times 10^{-9} \tan^3\!\beta
    \left( \frac{v}{m_{\rm SUSY}} \right).
\end{equation}
Interestingly, the predictions are small enough to evade the current 
experimental constraints of $O(10^{-6})$ but large enough to be tested 
at future experiments.  For $\tau \to e$ and $\tau \to \mu$ transitions, 
we obtain
\begin{equation}
  \frac{B(\tau \to e \gamma)}{B(\mu \to e \gamma)} 
  \sim 0.2 \left[ \frac{\epsilon_1^{(\bar{5})} \epsilon_3^{(\bar{5})}}
    {\epsilon_1^{(\bar{5})} \epsilon_2^{(\bar{5})}} \right]^2
  \sim 0.2,
\quad
  \frac{B(\tau \to \mu \gamma)}{B(\mu \to e \gamma)} 
  \sim 0.2 \left[ \frac{\epsilon_2^{(\bar{5})} \epsilon_3^{(\bar{5})}}
    {\epsilon_1^{(\bar{5})} \epsilon_2^{(\bar{5})}} \right]^2
  \sim 20,
\end{equation}
for $\tan \beta \simgt 10$, finding the same relations as 
Eq.~(\ref{eq:mu-tau}).

The largest contribution to the $K^0$-$\bar{K}^0$ mixing, $\Delta m_K$, 
comes from the mass insertion $(\delta^d_{12})_{RR}$:
\begin{equation}
  (\delta^d_{12})_{RR} 
  \simeq \epsilon_1^{(\bar{5})} \epsilon_2^{(\bar{5})} 
  \simeq 1 \times 10^{-5} \tan^2\!\beta.
\end{equation}
The experimental constraint of order $10^{-2}\!\sim\!10^{-1}$ can be 
easily satisfied unless $\tan\beta$ is extremely large.  Relations 
among various meson mixings are predicted to be:
\begin{equation}
  \frac{\Delta m_{B_d}}{\Delta m_K} \sim 1, \qquad
  \frac{\Delta m_{B_s}}{\Delta m_K} \sim 10^2, \qquad
  \frac{\Delta m_{D}}{\Delta m_K} \sim \frac{10^2}{\tan^4\!\beta}.
\end{equation}
With the experimental bound on $\Delta m_K$, it will be difficult to 
see deviations from the standard model predictions in $B$ and $D$ meson 
systems.

Finally, the gluino mediated $b \to s \gamma$ decay may occur through 
\begin{equation}
  (\delta^d_{23})_{RL}^{\rm eff} 
  \simeq \epsilon_2^{(\bar{5})} \epsilon_3^{(\bar{5})} 
    \frac{m_b \tan \beta }{m_{\rm SUSY}}
  \simeq 7 \times 10^{-7} \tan^3\!\beta
    \left( \frac{v}{m_{\rm SUSY}} \right).
\end{equation}
The experimental bound of order $10^{-2}$ is also satisfied.

\subsection{Impact on inflation}
\label{subsec:inflation}

In this subsection we make a comment on our assumption of large ``wavefunction 
factors.''  Instead of introducing these factors in the function ${\cal F}$, 
as was done in Eqs.~(\ref{eq:F}~--~\ref{eq:W}), we could introduce similar 
factors in the K\"ahler potential $K$ (i.e. large factors only in front 
of the non-holomorphic quadratic terms in $K$).  These two assumptions 
are physically distinct.  For example, the ``${\cal F}$-based'' case leads 
to (flavor universal) higher dimension operators in $K$, suppressed only 
by powers of the fundamental scale $M_*$.  In the ``$K$-based'' case, on 
the other hand, these operators receive additional suppressions due to 
$\epsilon$'s (in the basis where fields are canonically normalized). 
A nonvanishing $m_0^2$ of order $m_{3/2}^2$ is also automatically obtained 
in this case.  While the naive extra dimensional picture leads to the 
${\cal F}$-based form, we do not find anything particularly wrong for 
the $K$-based form from a purely phenomenological point of view.  (In 
particular, such an assumption is radiatively stable.) 

It is interesting to point out that if we adopt the $K$-based assumption 
and introduce a large ${\cal Z}$ factor for a field, then the field has 
an almost minimal K\"ahler potential.  This can provide a solution to the 
``$\eta$ problem'' of inflationary models, when applied to the inflaton 
field.  This is because, assuming that the superpotential is (effectively) 
linear in the inflaton field, as in the case of hybrid inflation models, 
the minimal K\"ahler potential can avoid a generation of unwanted 
supergravity-induced mass term, of order the Hubble parameter, for the 
inflaton field~\cite{Kumekawa:1994gx}.  It is interesting that another 
fine-tuning problem of supersymmetric models -- the $\eta$ problem -- 
might be connected to the naturalness problems we are addressing, i.e. 
those of electroweak symmetry breaking, dark matter, and supersymmetric 
flavor.

\subsection{Generalization of the model}
\label{subsec:gen}

The analyses so far have assumed unified relations on supersymmetry 
breaking parameters, such as universal gaugino masses and common wavefunction 
factors $\epsilon_{Q_i} \simeq \epsilon_{U_i} \simeq \epsilon_{E_i}$ 
and $\epsilon_{D_i} \simeq \epsilon_{L_i}$ for unified multiplets. 
In this subsection we discuss possible deviations from these assumptions. 

We first consider the case where the universality of the gaugino masses is 
relaxed.  We do not expect significant changes in Fig.~\ref{fig:region} in 
this case.  The allowed region for $m_0 = 0$ is mainly controlled by the 
right-handed slepton masses and the bino mass, which are determined only 
by $M_1$ and the one-loop induced $U(1)_Y$ $D$-term, and thus there is 
no significant effect by the non-universality.  The Higgs boson mass, which 
is important for both cases of vanishing and nonvanishing $m_0$, is mainly 
determined by the $A_t$ term and the top squark masses at low energies. 
These quantities can be significantly modified by changing the $M_3$ parameter, 
but its effects can be compensated by changing $A_t$, $m_{\tilde{Q}_3}^2$ 
and $m_{\tilde{U}_3}^2$ at the unification scale.  The dark matter abundance 
is again mainly determined by $M_1$, $\mu$, and $m_A$.  Therefore, as far 
as the allowed region is concerned, the non-universality is not so important. 
However, it may be important for fine-tuning in electroweak symmetry breaking. 
Lowering $M_3$ leads to a suppression of the mass scale for all the colored 
superparticles, especially for the top squarks, reducing the one-loop 
correction to the Higgs mass squared parameter, $m_{H_u}^2$.

Modifications of the relations among different ${\cal Z}_r$ factors 
may cause quite dramatic changes in the predictions of flavor changing 
processes.  However, the large branching ratio of $\mu \to e \gamma$ 
decay is quite generic, since 
\begin{equation}
  \max \left[ (\delta^l_{12})_{LR}, (\delta^l_{12})_{RL} \right] 
  \simgt 3 \times 10^{-5} \left( \frac{v}{m_{\rm SUSY}} \right),
\label{eq:meg_general-1}
\end{equation}
\begin{eqnarray}
  \max \left[ (\delta^l_{12})_{LR}^{\rm eff}, (\delta^l_{12})_{RL}^{\rm eff} \right]
  \simgt 1 \times 10^{-8} \tan^2\!\beta \left( \frac{v}{m_{\rm SUSY}} \right),
\label{eq:meg_general-2}
\end{eqnarray}
both of which result only from $\epsilon_{L_1} \epsilon_{E_1} \simeq (y_e)_{11} 
\simeq m_e / (v \cos \beta)$ and $\epsilon_{L_2} \epsilon_{E_2} \simeq 
(y_e)_{22} \simeq m_\mu / ( v \cos \beta)$.  Here, Eq.~(\ref{eq:meg_general-1}) 
is for the case where the superpotential couplings $XLHE$ are present with 
$O(1)$ coefficients, while Eq.~(\ref{eq:meg_general-2}) for the case where 
these couplings are suppressed.

\section{Conclusions}
\label{sec:concl}

For phenomenological analyses of supersymmetric models, the $\mu$ 
term is often taken not to be a fundamental input parameter but rather 
provided as a solution to the constraint equation for electroweak symmetry 
breaking.  This conventional approach could hide phenomenologically important 
parameter regions.  For example, in well-studied models such as minimal 
supergravity or gauge mediation models, large values are generally predicted 
for the $\mu$ parameter.  However, this is by no means a general prediction 
of supersymmetric models.  In fact, a large $\mu$ parameter is rather 
disfavored from a purely phenomenological point of view.

The large $\mu$ term in conventional models is caused by a large negative 
$m_{H_u}^2$ parameter at low energies, which requires a precise cancellation 
between $m_{H_u}^2$ and $\mu^2$ in reproducing the correct scale for 
electroweak symmetry breaking, $v \simeq 174~{\rm GeV}$.  This cancellation 
is the source of the supersymmetric fine-tuning problem.  Turning the 
argument around, once we assume that there is no such fine-tuning for 
some reason, the $\mu$ parameter should not be so large compared with 
the electroweak scale.  In fact, this is even true in models with extended 
Higgs sectors.  The effective $\mu$ parameter, which parametrizes the 
supersymmetric contribution to the Higgs potential (or the Higgsino 
mass), should not be large --- no matter what its origin is.

After realizing that a large $\mu$ parameter is obtained only as 
a consequence of fine-tuning, it is sensible to take $\mu$ as an input 
parameter and study phenomenology of weak scale supersymmetry with 
a small $\mu$ parameter.  The most striking effect is that the small 
$\mu$ term enhances the mixing between the Higgsino and the bino and 
significantly reduces the thermal relic abundance of the bino dark 
matter.  We have shown that, with a small $\mu$ term, it is indeed 
quite easy to realize the neutralino dark matter without living in special 
parameter regions, such as near the $A$-pole or coannihilation regions. 
Furthermore, in such a situation, the detection rate for the neutralino 
dark matter in direct detection experiments is significantly enhanced. 
In the case where the gaugino masses are universal at the unification 
scale, we have obtained an absolute lower bound on the spin-independent 
cross section, $\sigma_{\rm SI} \simgt 10^{-46}~{\rm cm}^2$, for 
electroweak fine-tuning no worse than $\approx 5\%$.

A possible realization of a small $\mu$ term is obtained by deviating 
from minimal supergravity by changing the $m_{H_u}^2$ parameter at the 
unification scale.  We have presented a simple model to realize this 
situation, which is achieved by placing the supersymmetry breaking 
sector at the same ``location'' as the electroweak symmetry breaking 
sector (the Higgs fields).  The hierarchy of the Yukawa couplings then 
implies that the third generation fields live ``close'' to the location 
of the Higgs fields, while the first two generations ``away'' from it. 
With this setup, the Higgs fields and the third generation sfermions feel 
supersymmetry breaking directly, while the first two generations only 
through renormalization group effects, which is desired for satisfying 
constraints from flavor changing processes.  We have found that such 
a pattern of supersymmetry breaking masses indeed leads to viable 
parameter regions, where all the experimental constraints are satisfied 
and the dark matter of the universe is explained by the thermal relic 
abundance of the lightest neutralino.  Making $A_t$ large at the 
unification scale facilitates to evade the Higgs boson mass bound with 
a low overall scale of supersymmetry breaking masses, reducing fine-tuning 
(equivalent to reducing the $\mu$ parameter).  Low-energy flavor 
violating processes are tightly related to the structure of the Yukawa 
couplings, so that their rates can be estimated.  We have found that 
they are consistent with the current experimental bounds, but some of 
them are close.  In particular, the $\mu \to e$ transition rates are 
predicted to be large, so that these processes should be discovered 
in near future experiments.

\section*{Acknowledgments}

The work of R.K. was supported by the U.S. Department of Energy under 
contract number DE-AC02-76SF00515.  The work of Y.N. was supported 
in part by the Director, Office of Science, Office of High Energy 
and Nuclear Physics, of the US Department of Energy under Contract 
DE-AC02-05CH11231, by the National Science Foundation under grant 
PHY-0403380, by a DOE Outstanding Junior Investigator award, and 
by an Alfred P. Sloan Research Fellowship.

\end{document}